\documentclass[twocolumn,apj]{aastex63}
\pdfoutput=1
\usepackage{amsmath,amssymb,gensymb,color,graphicx,array,multirow,soul}
\usepackage{natbib,hyperref,multirow,placeins,textcomp}
\usepackage[outline]{contour}
\usepackage[utf8]{inputenc}
\usepackage[T1]{fontenc}
\usepackage{booktabs}
\usepackage{microtype}
\usepackage{listings}

\DeclareFixedFont{\ttb}{T1}{txtt}{bx}{n}{9} 
\DeclareFixedFont{\ttm}{T1}{txtt}{m}{n}{9}  

\newcommand{\dfn}[1]{\textbf{#1}}

\newenvironment{closetabcols}[1][0.5mm]{\setlength{\tabcolsep}{#1}}{}
\newenvironment{closetabrows}[1][0.02]{\renewcommand{\arraystretch}{#1}}{}

\begin{document}

\title{The Atacama Cosmology Telescope: A search for Planet~9}

\author[0000-0002-4478-7111]{Sigurd~Naess}
\affil{Center for Computational Astrophysics, Flatiron Institute, New York, NY, USA 10010}
\author[0000-0002-1035-1854]{Simone~Aiola}
\affil{Center for Computational Astrophysics, Flatiron Institute, New York, NY, USA 10010}
\author[0000-0001-5846-0411]{Nick~Battaglia}
\affil{Department of Astronomy, Cornell University, Ithaca, NY 14853, USA}
\author[0000-0003-2358-9949]{Richard~J.~Bond}
\affil{Canadian Institute for Theoretical Astrophysics, 60 St. George Street, University of Toronto, Toronto, ON, M5S 3H8, Canada}
\author[0000-0003-0837-0068]{Erminia~Calabrese}
\affil{School of Physics and Astronomy, Cardiff University, The Parade, Cardiff, Wales, UK CF24 3AA}
\author[0000-0002-9113-7058]{Steve~K.~Choi}
\affil{Department of Physics, Cornell University, Ithaca, NY 14853, USA}
\affil{Department of Astronomy, Cornell University, Ithaca, NY 14853, USA}
\author[0000-0002-6151-6292]{Nicholas~F.~Cothard}
\affil{Department of Physics, Cornell University, Ithaca, NY 14853, USA}
\author[0000-0002-1760-0868]{Mark~Halpern}
\affil{Department of Physics and Astronomy, University of British Columbia, Vancouver, BC, Canada V6T 1Z4}
\author[0000-0002-9539-0835]{J.~Colin~Hill}
\affil{Department of Physics, Columbia University, New York, NY, USA 10027}
\affil{Center for Computational Astrophysics, Flatiron Institute, New York, NY, USA 10010}
\author[0000-0003-0744-2808]{Brian~J.~Koopman}
\affil{Department of Physics, Yale University, New Haven, CT 06520, USA}
\author[0000-0002-3169-9761]{Mark~Devlin}
\affil{Department of Physics and Astronomy, University of Pennsylvania, 209 South 33rd Street, Philadelphia, PA, USA 19104}
\author[0000-0002-7245-4541]{Jeff~McMahon}
\affil{Kavli Institute for Cosmological Physics, University of Chicago, Chicago, IL 60637, USA}
\affil{Department of Astronomy and Astrophysics, University of Chicago, Chicago, IL 60637, USA}
\affil{Department of Physics, University of Chicago, Chicago, IL 60637, USA}
\affil{Enrico Fermi Institute, University of Chicago, Chicago, IL 60637, USA}
\author[0000-0002-1940-4289]{Simon Dicker}
\affil{Department of Physics and Astronomy, University of Pennsylvania, 209 South 33rd Street, Philadelphia, PA, USA 19104}
\author[0000-0003-2856-2382]{Adriaan~J.~Duivenvoorden}
\affil{Joseph Henry Laboratories of Physics, Jadwin Hall, Princeton University, Princeton, NJ, USA 08544}
\author[0000-0002-7450-2586]{Jo~Dunkley}
\affil{Joseph Henry Laboratories of Physics, Jadwin Hall, Princeton University, Princeton, NJ, USA 08544}
\affil{Department of Astrophysical Sciences, Peyton Hall, Princeton University, Princeton, NJ, USA 08544}
\author[0000-0003-2410-0922]{Valentina Fanfani}
\affil{Department of Physics, University of Milano-Bicocca, Piazza della Scienza, 3 - 20126 Milano (MI), Italy}
\author[0000-0003-4992-7854]{Simone~Ferraro}
\affil{Lawrence Berkeley National Laboratory, Berkeley, California 94720, USA}
\affil{Berkeley Center for Cosmological Physics, University of California, Berkeley, California 94720, USA}
\author[0000-0001-9731-3617]{Patricio~A.~Gallardo}
\affil{Department of Physics, Cornell University, Ithaca, NY 14853, USA}
\author[0000-0002-1697-3080]{Yilun~Guan}
\affil{Department of Physics and Astronomy, University of Pittsburgh, Pittsburgh, PA, USA 15260}
\author[0000-0001-5649-3551]{Dongwon~Han}
\affil{Physics and Astronomy Department, Stony Brook University, Stony Brook, NY 11794}
\author[0000-0002-2408-9201]{Matthew~Hasselfield}
\affil{Center for Computational Astrophysics, Flatiron Institute, New York, NY, USA 10010}
\author[0000-0003-1690-6678]{Adam~D.~Hincks}
\affil{Department of Astronomy and Astrophysics, University of Toronto, 50 St. George Street, Toronto, ON M5S 3H4, Canada}
\author[0000-0001-7109-0099]{Kevin~Huffenberger}
\affil{Department of Physics, Florida State University, Tallahassee FL, USA 32306}
\author[0000-0002-3734-331X]{Arthur~B.~Kosowsky}
\affil{Department of Physics and Astronomy, University of Pittsburgh, Pittsburgh, PA, USA 15260}
\author[0000-0002-6849-4217]{Thibaut~Louis}
\affil{Universit\'e Paris-Saclay, CNRS/IN2P3, IJCLab, 91405 Orsay, France}
\author{Amanda Macinnis}
\affil{Physics and Astronomy Department, Stony Brook University, Stony Brook, NY 11794}
\author[0000-0001-6740-5350]{Mathew~S.~Madhavacheril}
\affil{Perimeter Institute for Theoretical Physics, 31 Caroline Street N, Waterloo ON N2L 2Y5 Canada}
\author[0000-0002-8307-5088]{Federico Nati}
\affil{Department of Physics, University of Milano-Bicocca, Piazza della Scienza, 3 - 20126 Milano (MI), Italy}
\author[0000-0001-7125-3580]{Michael~D.~Niemack}
\affil{Department of Physics, Cornell University, Ithaca, NY 14853, USA}
\affil{Department of Astronomy, Cornell University, Ithaca, NY 14853, USA}
\affil{Kavli Institute at Cornell for Nanoscale Science, Cornell University, Ithaca, NY 14853, USA}
\author[0000-0002-9828-3525]{Lyman~Page}
\affil{Joseph Henry Laboratories of Physics, Jadwin Hall, Princeton University, Princeton, NJ, USA 08544}
\author[0000-0003-4006-1134]{Maria~Salatino}
\affil{Physics Department, Stanford University Kavli Institute for Particle Astrophysics and Cosmology (KIPAC) Stanford, California CA}
\author[0000-0002-4619-8927]{Emmanuel~Schaan}
\affil{Lawrence Berkeley National Laboratory, Berkeley, California 94720, USA}
\affil{Berkeley Center for Cosmological Physics, University of California, Berkeley, California 94720, USA}
\author[0000-0003-1842-8104]{John Orlowski-Scherer}
\affil{Department of Physics and Astronomy, University of Pennsylvania, 209 South 33rd Street, Philadelphia, PA, USA 19104}
\author[0000-0002-0512-1042]{Alessandro~Schillaci}
\affil{Department of Physics, California Institute of Technology, Pasadena, California 91125, USA}
\author{Benjamin~Schmitt}
\affil{Harvard-Smithsonian Center for Astrophysics}
\author[0000-0002-9674-4527]{Neelima~Sehgal}
\affil{Physics and Astronomy Department, Stony Brook University, Stony Brook, NY 11794}
\author[0000-0002-8149-1352]{Crist\'obal~Sif\'on}
\affil{Instituto de F\'isica, Pontificia Universidad Cat\'olica de Valpara\'iso, Casilla 4059, Valpara\'iso, Chile}
\author[0000-0002-7020-7301]{Suzanne~Staggs}
\affil{Joseph Henry Laboratories of Physics, Jadwin Hall, Princeton University, Princeton, NJ, USA 08544}
\author{Alexander~Van~Engelen}
\affil{School of Earth and Space Exploration, Arizona State University, Tempe, AZ, USA 85287}
\author[0000-0002-7567-4451]{Edward~J.~Wollack}
\affil{NASA/Goddard Space Flight Center, Greenbelt, MD, USA 20771}

\keywords{}

\defcitealias{p9-proposal}{B16}
\defcitealias{p9-hypothesis}{B19}
\defcitealias{p9-atmosphere-2016}{F16}

\begin{abstract}
	We use Atacama Cosmology Telescope (ACT) observations at 98 GHz (2015--2019), 150 GHz (2013--2019) and 229 GHz (2017--2019) to perform a blind shift-and-stack search for Planet~9. The search explores distances from 300 AU
	to 2000 AU and velocities up to 6.3 arcmin per year, depending on the distance ($r$). For
	a 5 Earth-mass Planet~9 the detection limit varies from 325 AU to 625 AU, depending on the sky location.
	For a 10 Earth-mass planet the corresponding range is 425 AU to 775 AU.
	The search covers the whole 18\,000 square degrees of the ACT survey.
	No significant detections are found, which is used to place limits on the mm-wave flux density of Planet~9
	over much of its orbit. Overall we eliminate roughly 17\% and 9\% of the parameter space for
	a 5 and 10 Earth-mass Planet~9 respectively. We also provide a list of the 10 strongest candidates from the search
	for possible follow-up. More generally, we exclude (at 95\% confidence) the presence of an unknown Solar
	system object within our survey area brighter than 4--12 mJy (depending on position) at 150 GHz with current
	distance $300 \text{ AU} < r < 600 \text{ AU}$ and heliocentric angular velocity
	$1.5'/\text{yr} < v \cdot \frac{500 \text{ AU}}{r} < 2.3'\text{yr}$, corresponding to low-to-moderate eccentricities.
	These limits worsen gradually beyond 600 AU, reaching 5--15 mJy by 1500 AU.
	\vspace*{5mm}
\end{abstract}

\section{Introduction}
The existence of ``Planet~9'', a large (mass $M \sim 5-10$ $M_\Earth$) and very distant
(semi-major axis $a \sim 400-800$ AU)
new planet in the solar system, has recently been proposed as an explanation for the
observed clustering of orbits of the highest-perihelion objects in the detached Kuiper
belt \citep{p9-proposal,p9-hypothesis} (hereafter B16 and B19). While the reality of
this clustering is unclear because of the presence of large observational biases
\citep{shankman-tno-bias-2017,des-tno-isotropy,no-p9-napier,brown-p9-response},
the hypothesis has still gathered considerable interest.

Most new solar system objects are discovered in optical surveys via their reflected
sunlight. At these wavelengths, Planet~9 would appear as a magnitude 19--24 object (depending
on the size and distance and assuming an albedo between 0.4 and 1) \citepalias[page 61]{p9-hypothesis}:
quite faint due to the $1/r^4$ dependence of reflected sunlight,\footnote{
	Here $r$ is the object's current distance from the Sun. Technically the expression
	should be $1/(r^2r_\Earth^2)$ where $r_\Earth$ is the distance from the Earth, but in the
	outer solar system $r \approx r_\Earth$.} but still detectable by optical surveys like
the Dark Energy Survey (DES), the Hyper-Suprime Cam survey (HSC) or the Legacy Survey of
Space and Time (LSST).

The steep fall-off of flux density with distance can be circumvented by observing at longer
wavelengths, where thermal emission dominates. The heat budget of large objects far
from the Sun is dominated by their gravitational contraction and residual heat of
formation, resulting in a temperature that is approximately independent of their
distance from the Sun. This leads to a much gentler $1/r^2$ dependence. For sufficiently
large distances this can partially compensate for, or even overcome, the
resolution advantage enjoyed by optical surveys compared to those at mm or sub-mm wavelengths.
Indeed, the best current limits on the existence of Saturn- or Jupiter-size
trans-Neptunian objects (TNOs) is the Wide-Infrared Survey Explorer (WISE),
which observes in the 2.8--26 $\micro$m (11--110 THz) range. WISE has excluded
the existence of a Saturn-size planet out to 28\,000 AU, and a Jupiter-size one
out to 82\,000 AU \citep{wise-planet-lims}. Sadly, these limits degrade very
quickly with mass, partially because of a decrease in surface area, but more importantly
because lower-mass planets cool down more quickly. For sufficiently low masses,
the majority of the thermal emission would fall outside the WISE frequency range,
and this is expected to be the case for typical atmospheric models (see section~\ref{sec:p9-phys}).
However, emission predictions in the 3-5 $\micro$m window are extremely model
dependent, varying by four orders of magnitude. The brightest of these could be detectable by WISE.
\citet{p9-wise-2018} report a non-detection of Planet~9 in WISE's
3.6 $\micro$m W1 band, limiting its W1 magnitude to $>16.7$ (flux density $<65 \text{ } \micro$Jy) at
90\% confidence. For the most optimistic atmospheric models, this excludes a $10 M_\Earth$ Planet~9
up to 900 AU, but for more typical cases
WISE would not be sensitive to Planet~9's thermal radiation, motivating a search at
lower frequencies.

Soon after Planet~9 was first proposed, \citet{p9-cmb-cowan} (and later
\citet{p9-cmb-baxter})
suggested a search using Cosmic Microwave Background (CMB) telescopes
operating in the 1--3 mm range. The only current CMB survey telescopes with
high enough resolution to have any hope of detecting a faint, unresolved
object like Planet 9 are the South Pole Telescope (SPT) \citep{carlstrom/etal/2011}
and the Atacama Cosmology Telesope (ACT) \citep{fowler/etal:2007,thornton/2016},
and of these only ACT covers the low ecliptic latitudes where Planet~9 might
lurk.

ACT is a 6-m mm-wave telescope located
at 5190 m altitude on Cerro Toco in the northern Chilean Andes.
ACT began observations in 2008, and has been upgraded several times to
add polarization support and increase its sensitivity and frequency coverage.
ACT is currently surveying 18\,000 square degrees of the sky in five broad bands
roughly centered on 27 GHz, 39 GHz, 98 GHz, 150 GHz and 229 GHz, though the first two were
added too recently to be available for this analysis. We label these
bands f030, f040, f090, f150 and f220 respectively.

The primary goal of the survey is
to map the CMB, but the telescope's relatively
high angular resolution of 2.05/1.40/0.98 arcminutes full-width-half-max (FHWM)
in the f090/f150/f220 bands respectively makes it capable of a large set of
other science goals, including searches for galaxy clusters, active galactic nuclei
and transients. We here report on a search for Planet~9 using 7 years of ACT data
collected from 2013 to 2019.

\section{Planet~9 in the ACT bands}
\label{sec:p9-phys}
\citet{p9-atmosphere-2016} (henceforth F16) investigated the radius, temperature and luminosity
of Planet~9, and found that the Sun had a minimal impact on its heat budget, and hence
its physical properties do not depend on the planet's distance from the Sun.
They do, however, depend considerably on both its mass and internal composition,
for which \citetalias{p9-atmosphere-2016} build several models.

Their nominal scenario has a H/He envelope making up 10\% of the planet's mass,
with the remainder being mostly a 2:1 mix of ice and rock. For this composition
they find that the most favored $5 M_\Earth$ scenario of \citetalias{p9-hypothesis} results
in a radius of $2.94 R_\Earth$, a temperature
of 42.2 K and a featureless blackbody spectrum below 8 THz.\footnote{Note:
\citetalias{p9-atmosphere-2016} cautions that while their framework fits
Neptune well, it overestimates Uranus' temperature, and they cannot
exclude that this could be the case for Planet~9 too.} For a fiducial distance
of 500 AU, this results in a flux density of 2.3 mJy, 5.3 mJy and 11 mJy in the three ACT bandpasses
f090, f150 and f220. For the $10 M_\Earth$ scenario, which is near the upper end of the possible
mass range, the corresponding numbers are $R = 3.46 R_\Earth$, $T = 48.3 \text{K}$ and
a flux density of 3.7/8.5/18 mJy at f090/f150/f220. These numbers vary by 10--50\%
depending on the composition -- see Table~\ref{tab:p9-props}.\footnote{
	This ignores the small loss of flux density that comes from the planet blocking
the 2.725 K CMB monopole. This leads to a 2.6/1.4/0.6\% loss of flux density
at f090/f150/f220, which is negligible compared to the uncertainty on Planet~9's physical
properties.} Depending on Planet~9's
exact orbit, its current distance could vary from about 300 AU to 1200 AU, but due
to the radius and temperature being independent of the distance from the Sun, this
simply rescales the flux densities as $1/r^2$.

The expected distance to Planet~9 is
correlated with its mass, since a more massive planet has to be further away to avoid
having too large of an effect on the orbits of other trans-Neptunian objects. A $5 M_\Earth$
Planet~9 would have an expected semi-major axis $a \sim 500$ AU and an eccentricity
of $0.1 \lessapprox e \lessapprox 0.3$, while at $10 M_\Earth$ the
best-fit semi-major axis and eccentricity are $a \sim 700$ AU and $0.3 \lessapprox e \lessapprox 0.5$ \citepalias[fig.~15]{p9-hypothesis}. At frequencies
$<2.5 \text{ THz}$, this increased distance mostly cancels the
increased luminosity of a more massive planet, making ACT's prospect for detecting
an object like Planet~9 only moderately sensitive to its mass.\footnote{This is in contrast to
$2.5 \text{ THz} < \nu < 20 \text{ THz}$ where small changes in mass lead to big changes
in detectability because of the steep fall of the blackbody spectrum here, and
$\nu > 20 \text{ THz}$ where the $1/r^4$ dependence of reflected sunlight makes a
smaller, closer planet much easier to detect.} The planet's inclination is
predicted to be moderate, $i<30^\circ$, with $i \approx 20^\circ$ preferred.

To see if ACT has any chance of detecting this signal, let us compare it to
ACT's sensitivity to stationary point sources. This varies by position in the map
but the 10--90\% quantile range is about 1--2 mJy at f090 and f150, and 4--8 mJy at f220.\footnote{For
comparison, the same quantile range for Planck 143 GHz is 29--41 mJy.}
Hence, if Planet~9 were stationary at 500 AU, we could expect
to detect it at $2.3-11 \sigma$ for the $5 M_\Earth$ case when combining the three ACT bands.
This is not high enough to guarantee a discovery, especially considering that Planet~9
could be at a larger distance than 500 AU, but it's high enough that a search is
worthwhile.

\begin{figure*}[htbp]
	\centering
	\hspace*{-4mm}\includegraphics[width=0.9\textwidth]{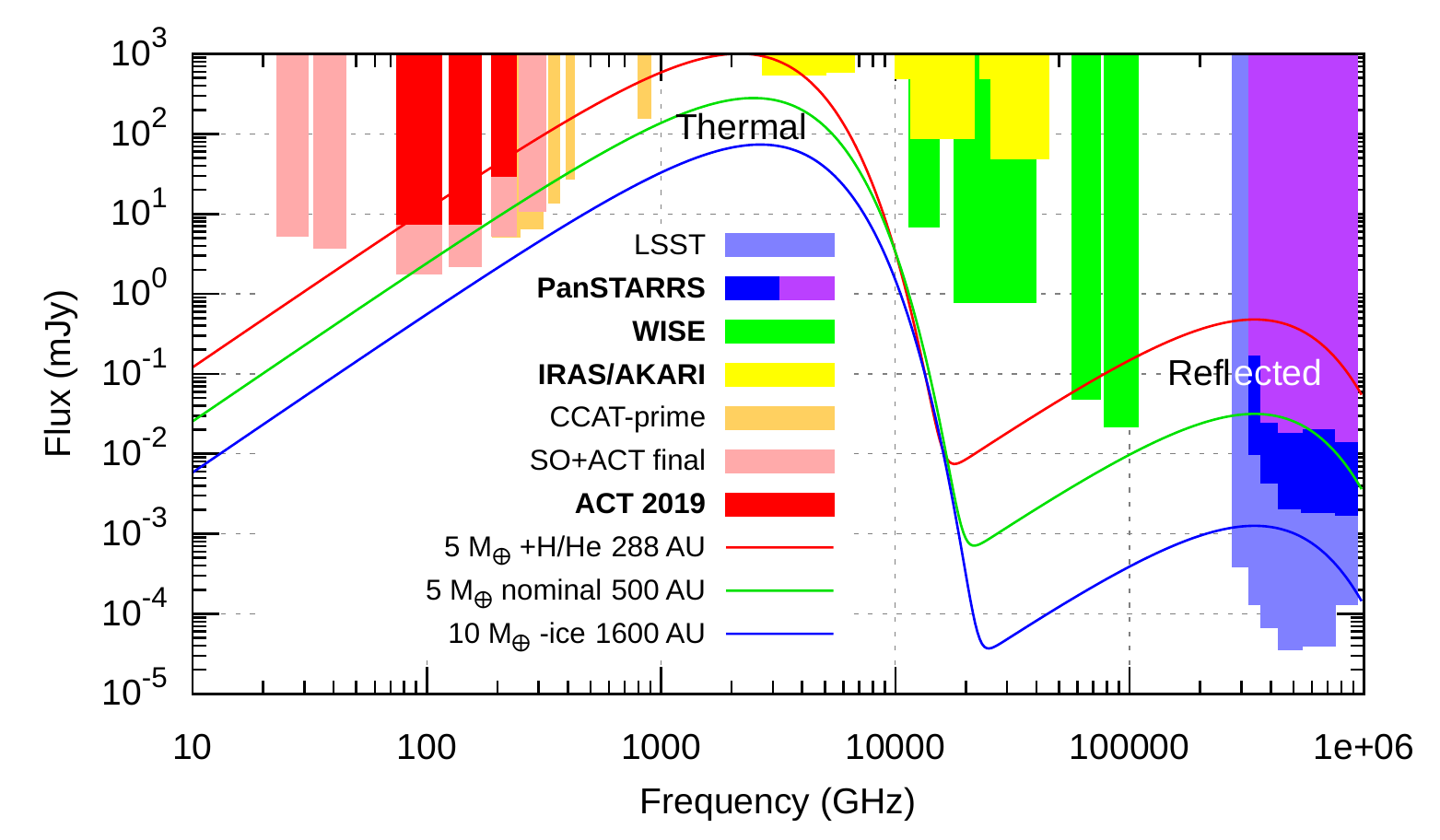}
	\caption{The potential Planet~9 spectra compared to the $5\sigma$ detection limit
	of current and upcoming wide-area surveys. Red curve: High-brightness scenario: a $5 M_\Earth$ Planet~9 with
	a heavy H/He envelope at a perihelion of 288 AU. Green curve: A more moderate
	scenario with a light H/He envelope and a 2:1 ice:rock ratio at 500 AU, still with $5 M_\Earth$. Blue curve:
	Low-brightness scenario: A $10 M_\Earth$ Planet~9 with a light H/He envelope and a 1:4 ice:rock
	ratio at an aphelion of 1160 AU. All scenarios assume unit emissivity black-body spectra,
	but differ in the planet radius and release and transport of internal gravitational energy;
	see \citetalias{p9-atmosphere-2016} Table 1.
	ACT 2019 is the data-set used in this paper, while SO+ACT final is the expected combined
	Simons Observatory \citep{so_science} + ACT data after both surveys finish. The others are CCAT-prime \citep{ccat-sens-2019},
	IRAS \citep{iras-psc}, AKARI \citep{akari-nir,akari-fir}, WISE \citep{allwise,unwise-catalog},
	Pan-STARRS \citep{panstarrs-survey-2016} and LSST \citep{lsst-overview}. Future surveys
	are shown with a thinner font and less intense color in the legend.
	For Pan-STARRS both the full depth (blue) and the transient search depth (purple) \citep{panstarrs-trans-surv} are shown.
	The double blackbody approximation used here may be inaccurate in the range 8-400 THz because of
	atmospheric features \citepalias[Fig.~1]{p9-atmosphere-2016}. The sensitivities shown are
	typical values for each survey, and do not attempt to compensate for the surveys' different
	sky coverage. Depth variations inside each survey and the effect of the large parameter space
	of a blind search for a moving object are ignored.
	}
	\label{fig:specs}
\end{figure*}

Figure~\ref{fig:specs} compares the brightest/medium/faintest expected
Planet~9 spectra (as inferred from the range of possible orbits from \citetalias{p9-hypothesis}
and of physical properties from \citetalias{p9-atmosphere-2016}) to the
sensitivity of ACT and other current and future
wide-area surveys. Despite WISE's impressive bounds on
Saturn- and Jupiter-size TNOs, it is not very sensitive to smaller,
and therefore colder, objects like Planet~9. The most sensitive current data set that covers
most/all of Planet~9's orbit is therefore Pan-STARRS. At its full depth of about magnitude
23, Pan-STARRS has a flux density limit of roughly 2 $\micro$Jy, but this degrades to around
20 $\micro$Jy (mag. $\approx 21$) if the search is limited to the depth of the
Pan-STARRS transient search (\citealp{panstarrs-trans-surv}, \citetalias{p9-hypothesis}).
Both WISE and Pan-STARRS have reduced sensitivity near the galactic plane because of
confusion. For the medium brightness case, ACT's typical depth could expect a
borderline detection, similar to the Pan-STARRS transient search and a bit better
than WISE.

\begin{table*}
	\hspace*{-20mm}
	\centering
	\begin{tabular}{ccccccccr}
		Mass & Radius & Temperature & Band & Flux @ 500 AU & ACT Depth & FWHM & Freq. \\
		$\text{M}_e$ &  $\text{R}_e$ &  K & - &  mJy &  mJy & arcmin & GHz \\
		\hline
		\multirow{3}{*}{\large 5} &
		\multirow{3}{*}{\large 4.12/2.94/2.71} &
		\multirow{3}{*}{\large 36.7/42.2/38.9} &
		f090 & 3.9/2.3/1.8 & 1.0--2.1 & 2.05 & 98 \\
		& & &
		f150 & 8.9/5.3/4.1 & 1.0--2.2 & 1.40 & 150 \\
		& & &
		f220 & 18/11/8.5 & 4.1--8.4 & 0.98 & 229 \\
		\hline
		\multirow{3}{*}{\large 10} &
		\multirow{3}{*}{\large 5.09/3.46/3.16} &
		\multirow{3}{*}{\large 40.3/48.3/45.1} &
		f090 & 6.6/3.7/2.9 & 1.0--2.1 & 2.05 & 98 \\
		& & &
		f150 & 15/8.5/6.6 & 1.0--2.2 & 1.40 & 150 \\
		& & &
		f220 & 31/18/14 & 4.1--8.4 & 0.98 & 229 \\
		\hline
	\end{tabular}
	\caption{
		Potential radii and temperatures for a $5 M_\Earth$ and $10 M_\Earth$ Planet~9 from
		\citetalias{p9-atmosphere-2016}.
		The three slash-separated entries correspond to three planet types
		described in their Table~1. The central one is the nominal
		case with a 2:1 ice:rock core surrounded by an H/He envelope.
		The leftmost entries are for a case with a larger H/He envelope and the
		rightmost entries are for the ice-poor case, for which the core is 1:2 ice:rock by mass.
		The corresponding flux density in the three ACT frequency bands
		for these cases is given in the flux density column, and compared to the ACT
		point source sensitivity, which is about 1--2 mJy as seen in the sixth column.
		The first/last number
		in the range of ACT depth is the 10\%/90\% quantile over the 18,000 square degrees
		that ACT covers. The maps are
		deep enough compared to the expected Planet~9 flux density that a search is worthwhile.
		Also shown are the ACT beam size and central frequency in each band.
	}
	\label{tab:p9-props}
\end{table*}

\section{The ACT data sets}
The data sets used in this analysis are identical to those used in \citet{act-coadd-2020}, except for the inclusion
of one more season of data (2019), and the exclusion of the Planck and ACT
MBAC data sets because of their low resolution and low sky coverage respectively.
This represents 7 years and 140 TB of data,
of which 81\%  was collected since the AdvACT camera \citep{advact-MF,advact-HF} became
operational in 2017 (i.e., after ACT Data Release 4).\footnote{Split by frequency, that's 37/72/17 TB at f090/f150/f220,
of which 93\%/71\%/100\% was collected since 2017.}
See Appendix~\ref{sec:data-appendix} for details.

\section{Search methodology}
\subsection{Blink comparison won't work}
The most common way to discover solar system objects is to look for objects
that have moved between two different exposures of the same patch of sky.
This method is fast, but is limited by the depth of each image, since the
object needs to be independently detected in both. This depth can be improved
with longer exposures, but this is limited by the angular velocity of the
object itself. Integrating longer than the time it takes the object to move
by the size of the beam will just smear it out without any further gains in S/N.
This is the regime ACT is in for Planet~9.

For ACT sky coverage and sensitivity, it would take 3--4 years of observations
just to have a chance of detecting a Planet~9-like object that was not moving in the sky.
By Kepler's laws, a planet with semi-major axis $a$, eccentricity $e$ and current solar distance $r$
will have a Sun-centered angular speed of
\begin{align}
	v &= 1.932'/\text{yr} \cdot \sqrt{\frac{a}{\text{500 AU}}} \left(\frac{r}{\text{500 AU}}\right)^{-2}\sqrt{1-e^2}. \label{eq:vorb}
\end{align}
At the same time, the Earth's orbit sweeps out a yearly parallax ellipse with a semi-major axis of
\begin{align}
	\theta_\pi &= \frac{\text{AU}}{r} = 6.875' \cdot \frac{500 \text{AU}}{r} \label{eq:parallax1}
\end{align}
corresponding to a maximum angular speed of
\begin{align}
	v_\pi &= \frac{2\pi\theta_\pi}{\textrm{yr}} = 43.20'/\text{yr} \cdot \frac{500 \text{AU}}{r}. \label{eq:parallax2}
\end{align}
For comparison, ACT has an angular resolution of 2.05/1.40/0.98' FWHM at f090/f150/f220 respectively.
To avoid excessive smearing we need $(v_\pi + v)\Delta t \approx
v_\pi \Delta t \ll \text{FWHM}$. For the smallest beam (f220) and a closest possible distance
of $r_\textrm{min} = 300$ AU, this gives us $\Delta t \ll 5 \text{ days}$. With 5 days of
integration time and the current ACT survey strategy the expected Planet~9 S/N would be $\sim 1$, more than 5 times too low for
a detection, or more than 25 times too low in terms of observing time!

\subsection{Shift and stack}
The smearing could be eliminated if one knew the orbit of the object one was looking for,
since that would allow one to shift each exposure to track the object as it moves across the sky.
In practice, while the Planet~9 hypothesis makes \emph{some} predictions about its orbit, they are far
too vague to allow for simple tracking like this. However, with enough computational
resources it is possible to loop through \emph{every} reasonable orbit, make a shifted stack
of individual short exposures using that orbit, and then look for objects in the resulting
image. This is the \emph{shift-and-stack} algorithm, and has been used to successfully detect
objects below the single-exposure sensitivity limit \citep{gladman-shift-stack-1998, holman-shift-stack-2018},

Planet~9's orbit is characterized by its 6 orbital elements: semi-major axis $a$,
eccentricity $e$, inclination $i$, longitude of ascending node $\Omega$, argument
of periapsis $\omega$ and true anomaly $\nu$. However, because of its large distance
and corresponding slow motion, it is sufficient for us to
consider its motion to be drifting linearly on the sky, modulated by parallax. This gives us the
following 5 free parameters:
\begin{enumerate}
	\item The heliocentric right ascension ($\alpha$) and declination ($\delta$) of the
		planet at a reference time $t_0$.
	\item The horizontal and vertical components of the heliocentric angular velocity
		$\vec v = [v_x, v_y]$.
		We define these in the local tangent plane, such that $\alpha_\text{obs} = \alpha + v_x (t-t_0) / \cos\delta_0$
		and $\delta_\text{obs} = \delta + v_y (t-t_0)$.
	\item The planet's current distance from the Sun, $r$, which we treat as constant in time.
\end{enumerate}

With these, the shift-and-stack algorithm takes the following general form:
\begin{enumerate}
	\item Split the data into chunks with duration $\Delta t$, and make a sky map of each.
	\item For each reasonable value of $r, v_x, v_y$, use these with the time $t$ of
		each map to shift them according to their constant heliocentric angular
		velocity and parallactic motion, and stack them to produce a combined map.
	\item Use a filter matched to the noise and signal properties to look for point sources in
		each combined map.
\end{enumerate}
We will go through the details of this process in the following sections.

\subsection{Mapping and the matched filter}
\subsubsection{The sky maps}
ACT observes the sky by sweeping backwards and forwards in azimuth while the
sky drifts past. As it does so, the temperature registered by the detectors
is read out hundreds of times per second, forming a vector of time-ordered-data $d$.
We model $d$ as
\begin{align}
	d &= Pm + n,
\end{align}
where $m$ is the (beam-convolved, pixelated\footnote{We use $0.5'$ pixels
in a Plate Carreé projection in equatorial coordinates. This is later downsampled
to $1'$ pixels (see Section~\ref{sec:search-space}).}) sky in $\micro$K CMB temperature units,
$P$ is a response matrix that
encodes the telescope's pointing as a function of time, and $n$ is
instrumental and atmospheric noise which we model as Gaussian covariance $N$. The maximum-likelihood
estimate for $m$ given $d$ is
\begin{align}
	\hat m &= (P^TN^{-1}P)^{-1} P^T N^{-1} d \notag \\
	M &= (P^TN^{-1}P)^{-1} \label{eq:mapmaking}
\end{align}
Here $M$ is the noise covariance matrix of the estimator $\hat m$.

\subsubsection{The matched filter}
To look for point sources in $\hat m$ we start by assuming that
all sources are far enough apart that they can be considered in isolation.
Our data model for a map containing a single point source in some pixel $p$
is then
\begin{align}
	\hat m &= RQs_p + u
\end{align}
where $s_p$ is the point source flux density in pixel $p$ in mJy at a reference frequency $\nu_0 = 150 \text{ GHz}$ and
$Q_i = \delta_{ip}$ is a vector that's unity at the source location in pixel $p$ and zero
elsewhere. It takes us from just a single flux density value to a map with that value in a single
pixel.

$R = B g(\nu_0,\nu) f(\nu) A_p$ is a response matrix that takes us from that map to beam-convolved $\micro$K at the
observed frequency. Here $B$ is the instrument beam normalized to have a pixel-space integral of one,
$g(\nu_0,\nu)$ is the conversion factor from flux density at the reference frequency $\nu_0$
to the observed frequency $\nu$, $f(\nu)$ is the conversion from flux density in mJy to
beam-convolved peak height in $\micro$K,
and $A_p$ is pixel area in steradians. Since we expect Planet~9 to be a blackbody with temperature
$T \approx 40$~K, we have $g(\nu_0,\nu) = b(\nu,T)/b(\nu_0,T)$, where
\begin{align}
	b(\nu,T) &= \frac{2h\nu^3}{c^2}\frac{1}{\exp\left(\frac{h\nu}{k_B T}\right)-1}
\end{align}
is the
Planck law for surface brightness $b$; and \footnote{In the expression for $f(\nu)$, $A_b^{-1}$ converts from mJy to mJy/sr,
$10^{-23}$ converts from mJy/sr to $\micro$W/m$^2$/Hz/sr, and the rest is the derivative of the Planck law
evaluated at $T=T_\text{CMB}$, and converts to linearized CMB units in $\micro$K.}
\begin{align}
	f(\nu) &= \left(\frac{2x^4 k_B^3 T_\text{CMB}^2}{h^2c^2} \frac{1}{4\sinh(x/2)^2} 10^{23} A_b\right)^{-1} \\
	x &= \frac{h\nu}{k_B T_\text{CMB}}. \notag
\end{align}
Finally, $u$ is the noise in $\hat m$ and has a covariance matrix $U$. For the purposes of point source detection,
$u$ consists of everything in $\hat m$ that isn't the point source, which includes both the instrumental
and atmospheric noise described by the $M$ covariance matrix from before, but also the CMB,
Cosmic Infrared Background (CIB), galactic dust, etc.

Given this model for $\hat m$, the maximum-likelihood estimate for the point source flux density at
the reference frequency is
\begin{align}
	\hat s_p &= \rho_p/\kappa_p, & \rho_p &= Q^TR^TU^{-1} \hat m \notag \\
	\sigma_{\hat s_p} &= 1/\sqrt{\kappa_p}, & \kappa_p &= Q^TR^TU^{-1}RQ \label{eq:apix}
\end{align}
Here $\sigma_{\hat s_p}$ is the standard deviation of $\hat s_p$,
$\kappa_p$ corresponding inverse variance, and $\rho_p$
is the inverse variance weighted flux density.

So far we have only estimated the point source flux density in some pixel $p$. But since
we don't a priori know where on the sky the planet could be, we need to estimate
the flux density in every pixel, resulting in the flux density sky map $\hat s$ and corresponding
uncertainty $\sigma_{\hat s}$ given by:\footnote{
Since $Q/Q^T$ just picks out
an individual row of the quantitiy it's applied to, $\rho_p$ is just element $p$
of the vector $R^TU^{-1}\hat m$ and $\kappa_p$ is just element $p$ along
the diagonal of $R^TU^{-1}R$.}
\begin{align}
	\hat s &= \rho/\kappa, & \rho &= R^TU^{-1} \hat m \notag \\
	\sigma_{\hat s} &= 1/\sqrt{\kappa}, & \kappa &= \text{diag}(R^TU^{-1}R) \label{eq:avec}
\end{align}
where the division is done pixel by pixel. The corresponding S/N is
\begin{align}
	S/N &= \vec s/\sigma_{\vec s} = \rho/\sqrt{\kappa} = \frac{R^TU^{-1} \hat m}{\sqrt{\text{diag}(R^TU^{-1}R)}}
\end{align}
which we recognize as the \emph{matched filter} for $\hat m$. This S/N
map is what one would usually use for object detection, e.g. by identifying
peaks with $S/N > 5$. As we shall see in Section~\ref{sec:norm}
the shift and stack parameter search complicates this, but the general idea
stays the same.

\subsubsection{Stacking}
If we have multiple estimates $\{\vec s_i\}$ built from independent chunks of
data, such as the few-day chunks we will use in the shift-and-stack algorithm,
these combine straightforwardly:\footnote{Unlike the previous section, where
e.g. $\rho_p$ was the value in a single pixel, here each $\rho_i$ is a whole
map.}
\begin{align}
	\rho_\text{tot} &= \sum_i \rho_i, &
	\kappa_\text{tot} &= \sum_i \kappa_i \notag \\
	\vec s_\text{tot} &= \rho_\text{tot}/\kappa_\text{tot}, &
	\sigma_{\vec s_\text{tot}} &= 1/\sqrt{\kappa_\text{tot}} \label{eq:stack}
\end{align}
Sadly, the presence of the same CMB, CIB etc. in each chunk of data breaks the
assumption of independence that this expression builds on. It would be possible to
build a more complicated expression that takes this into account, but given the
computationally expensive parameter search we perform we need the
stacking operation to be as fast and simple as possible. Thankfully we can
eliminate these correlated components by simply subtracting the time-averaged
mean of the sky from each chunk of data.

\subsubsection{Mean sky subtraction}
\label{sec:skysub}
We can avoid the complications of the CMB, CIB etc. acting as correlated
noise common to the data chunks by subtracting a high-S/N estimate of the mean sky
from each chunk of data before mapping it. This eliminates any static part of
the sky such as the CMB, CIB, galactic emission, etc. (including any we
don't know about), and leaves only time-dependent signals such as the planet
we're looking for, as well as variable point sources (which can be masked) and
transients (which are rare enough that we can ignore them). The cost is a small
increase in the noise if the mean sky model isn't noise-free, and a partial
subtraction of the signal itself that must be estimated and corrected for.  For
this search we use the ACT+Planck combined maps described in
\citet{act-coadd-2020}, but extended to include the 2019 season of data.

Aside from letting us stack using equation~\ref{eq:stack}, mean sky subtraction
has the effect of removing all but the instrumental and atmospheric noise from
the individual sky maps, and hence the matched filter noise covariance matrix $U$
reduces to $M$. Inserting this into equation~\ref{eq:avec} we get:
\begin{align}
	\rho &= R^T\overbrace{P^T N^{-1}d}^{\text{rhs}} \label{eq:rho}
\end{align}
The part labeled ``rhs'' is a map that is much cheaper to compute than $\hat m$ because it
avoids the expensive inversion $(P^TN^{-1}P)^{-1} = M$ which must usually be done
using iterative methods like Conjugate Gradients\footnote{
	This time save comes at a small cost. By using one $N^{-1}$ when building
	the numerator of equation~\ref{eq:avec}, but effectively a slightly
	different one in the denominator because of the approximation we have to
	do for $\kappa$, $N^{-1}$ no longer cancels in the
	expectation value and we introduce a small bias. This would have been
	avoided if we had computed the full $\hat m$ and then applied the same
	approximate $M^{-1}$ ($U^{-1}$) both in the numerator and denominator,
but is ultimately corrected during debiasing (Section~\ref{sec:bias}).
}. That leaves us with
$\kappa$ which we approximate as
\begin{align}
	\kappa_i &= R_{ji}M^{-1}_{jk} R_{ki} \approx \alpha R_{ji}^2 w_j
\end{align}
where the map $w$ is an approximation pixel-diagonal of $M^{-1}$ built assuming
white (uncorrelated) noise and $\alpha$ is a factor that
compensates for the mean error we make by replacing $M^{-1}$ with $\vec w$.
We determine $\alpha$ by evaluating a few pixels of the exact $\kappa$.

\subsubsection{Ad-hoc filter}
Due to the time-domain noise model underestimating the amount of correlated noise
in the data, we applied an extra ad-hoc filter to the maps. This is described in
Appendix~\ref{sec:ad-hoc-appendix}, but has the effect of suppressing noise
for scales $\gtrsim 0.1^\circ$.

\subsubsection{Point source handling}
During map-making, any samples that were within 0.8 degrees of Venus, Mars, Jupiter, Saturn,
Uranus or Neptune were cut to avoid both the planets themselves and 0.1--1\%-level contamination
through the near
sidelobes. In addition, any sample within 3 arcminutes of the bright asteroids Vesta,
Pallas, Ceres, Iris, Eros, Hebe, Juno, Melpomene, Eunomia, Flora, Bamberga, Ganymed,
Metis, Nausikaa and Malasslia were cut.

To avoid false detections from variable point sources (e.g. blazars) we also cut point sources with a peak amplitude
of at least 500 $\micro$K out to the radius where the beam has damped them to 10 $\micro$K.
For daytime data, the peak amplitude threshold was reduced to 150 $\micro$K and the
cut area was broadened by $\pm 1'$ in azimuth and $-1'$ to $4'$ in elevation to account for
the harder-to-model daytime beam and pointing. 500 $\micro$K corresponds to about 49/37/23
mJy in the f090/f150/f220 bands, and with this 2770/3054/1640 point sources were cut in the night and
9252/7886/1713 in the day. Point sources fainter than this (but still with $S/N>10$), of
which there were 8868/5246/73 for the night-time and 2382/413/0 for daytime were individually
fit and subtracted from the time-ordered data.

\subsubsection{Dust masking}
In theory all galactic dust should be canceled by the
mean sky subtraction, since this represents length scales too
large to evolve over the course of our observations. However,
in practice small time-variable errors in our detector calibration
can make the dust appear to fluctuate slightly in brightness.
For sufficiently bright regions of dust these fluctuations become
big enough
to induce a large number of false positives in the search. Ideally
we would use the dust signal itself to calibrate the detectors in
these regions, but for now we simply mask them.

We built a dust mask by high-pass filtering the Planck PR2 545 GHz map
with the Butterworth filter $\beta(\ell, 1500, -5)$ (see Appendix~\ref{sec:ad-hoc-appendix}),
selecting the
7\% brightest pixels of the absolute value of the result, and growing
the result by smoothing it with a Gaussian beam with $\sigma = 7.2$ arcmin
and masking areas with value $>0.5$.
This mask was applied to each $\rho,\kappa$ map. We found that
the edges of the mask introduce some artifacts during the shift-and-stack
search, so we additionally applied a 20 arcmin larger mask before the
final object detection step.

\subsection{Search space}
\label{sec:search-space}
\definecolor{cr1}{RGB}{0,50,255}
\definecolor{cr2}{RGB}{0,100,255}
\definecolor{cr3}{RGB}{0,149,255}
\definecolor{cr4}{RGB}{0,199,255}
\definecolor{cr5}{RGB}{0,212,175}
\definecolor{cr6}{RGB}{0,207,58}
\definecolor{cr7}{RGB}{60,198,0}
\definecolor{cr8}{RGB}{179,187,0}
\definecolor{cr9}{RGB}{255,162,0}
\definecolor{cr10}{RGB}{255,113,0}
\definecolor{cr11}{RGB}{239,67,0}
\definecolor{cr12}{RGB}{169,34,0}
\definecolor{cr13}{RGB}{100,0,0}
\definecolor{cr14}{RGB}{92,0,0}
\begin{figure*}[htbp]
	\centering
	\hspace*{-14mm}\begin{tabular}{>{\centering\arraybackslash}m{2mm}>{\centering\arraybackslash}m{5.6cm}>{\centering\arraybackslash}m{5.5cm}>{\centering\arraybackslash}m{5.6cm}}
		&
		\textbf{\textcolor{cr1}{300}}, \textbf{\textcolor{cr4}{375}}, \textbf{\textcolor{cr7}{500}}, \textbf{\textcolor{cr10}{750}} and \textbf{\textcolor{cr13}{1500}} AU &
		\textbf{\textcolor{cr2}{321}}, \textbf{\textcolor{cr5}{409}}, \textbf{\textcolor{cr8}{563}}, \textbf{\textcolor{cr11}{900}} and \textbf{\textcolor{cr14}{2000}} AU &
		\textbf{\textcolor{cr3}{346}}, \textbf{\textcolor{cr6}{450}}, \textbf{\textcolor{cr9}{643}} and \textbf{\textcolor{cr12}{1125}} AU \\
		\rotatebox[origin=c]{90}{$v_y$ ($'/\text{yr}$)\hspace*{-55mm}} &
		\includegraphics[width=0.31\textwidth]{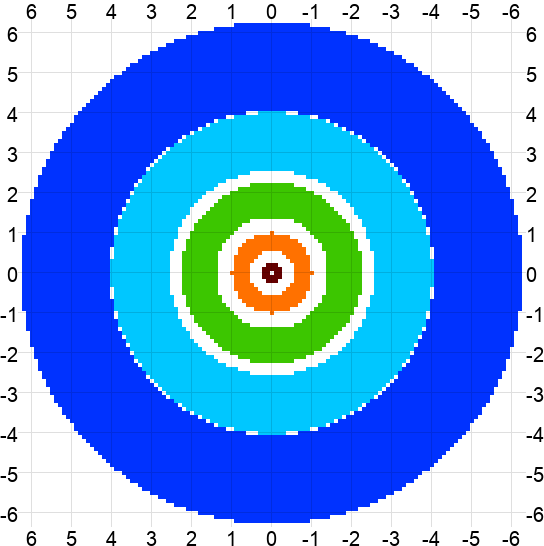} &
		\includegraphics[width=0.31\textwidth]{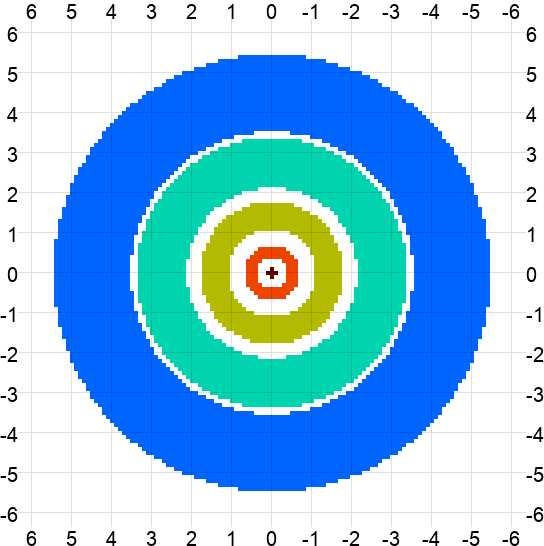} &
		\includegraphics[width=0.31\textwidth]{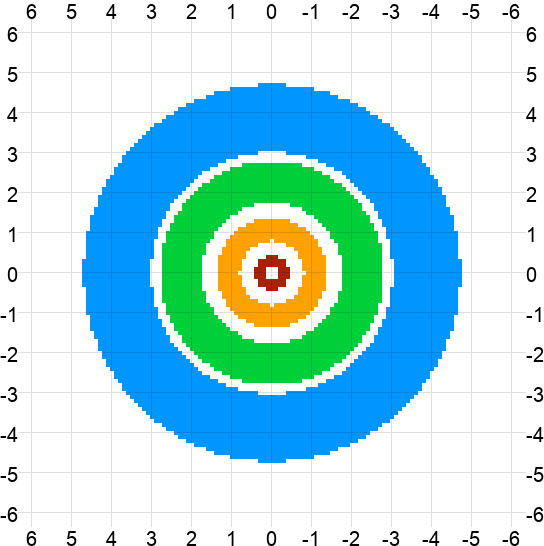} \\
		& $v_x$ ($'/\text{yr}$) & $v_x$ ($'/\text{yr}$) & $v_x$ ($'/\text{yr}$)
	\end{tabular}
	\caption{
		Illustration of the $r,v_x,v_y$ search space used in the Planet~9
		search. The horizontal and vertical axis shows $v_x$ and $v_y$
		respectively, the components of the heliocentric angular velocity,
		in units of $'$/yr (arcmin per year). Each colored
		region corresponds to the velocities that were explored for a
		given solar distance $r$. From blue to
		red these are 300, 321, 346, 375, 409, 450, 500, 563, 643, 750,
		900, 1125, 1500 and 2000 AU. The regions are spread over three
		sub-figures to avoid overlaps. The velocity grid used in the
		search had a 0.1$'$/yr resolution, resulting in a total parameter
		volume of 25\,837 cells.
	}
	\label{fig:search}
\end{figure*}

\begin{table}
	\centering
	\begin{closetabcols}
	\begin{tabular}{r!{\hspace*{2mm}}r!{-}r!{\hspace*{2mm}}r!{-}r!{\hspace*{2mm}}r!{-}r!{\hspace*{2mm}}r!{-}r!{\hspace*{2mm}}r!{-}r}
		$M$ & \multicolumn{2}{c}{$a$ (AU)} & \multicolumn{2}{c}{$e$} & \multicolumn{2}{c}{$q$ (AU)} & \multicolumn{2}{c}{$Q$ (AU)} & \multicolumn{2}{c}{$v_\text{ref}$ ($'$/yr)} \\
\hline
  5 &  350 & 450 & 0.10& 0.20&   280 & 405 &  385&  540&  1.58& 1.82 \\
  5 &  450 & 550 & 0.20& 0.30&   315 & 440 &  540&  715&  1.75& 1.99 \\
 10 &  650 & 750 & 0.30& 0.40&   390 & 525 &  845& 1050&  2.02& 2.26 \\
 10 &  750 & 850 & 0.40& 0.50&   375 & 510 & 1050& 1275&  2.05& 2.31
	\end{tabular}
	\end{closetabcols}
	\caption{
		Prior parameter ranges from Figure~15 of \citetalias{p9-hypothesis}.
		This is based on the best-fit parameter points for the $5 M_\Earth$ and $10 M_\Earth$
		scenarios, to which uncertainty ranges of $a \pm 50$ AU and
		$e \pm 0.05$ were added based on the resolution of the grid they
		used in their investigation. The planet mass, M, is given in Earth masses,
		e is the eccentricity, q is the perihelion distance and Q is the aphelion
		distance, both in AU.
		$v_\text{ref}$ is the planet's hypothetical speed at a reference location of $r_\text{ref}=500$ AU.
		The planet's actual speed will be $v = v_\text{ref} \frac{r_\text{ref}^2}{r^2}$.
	}
	\label{tab:prior}
\end{table}
The distance to and velocity of Planet~9 are relatively poorly determined,
but we can infer rough limits on the acceptable fit from Figure~15 of \citetalias{p9-hypothesis},
as shown in Table~\ref{tab:prior}. We see that Planet~9's current distance is limited to
$300 \text{AU} \lesssim r \lesssim 1300 \text{AU}$. Equation~\ref{eq:vorb} for the heliocentric
angular velocity of the planet can be re-expressed as
\begin{align}
	v &= v_\text{ref} \Big(\frac{r}{500 \text{AU}}\Big)^{-2} \\
	v_\text{ref} &= 1.93'/\text{yr} \cdot \sqrt{a(1-e^2)/(500 \text{ AU})}
\end{align}
and from Table~\ref{tab:prior} we see that $v_\text{ref}$ is in the range
1.6 to 2.3 $'$/yr for all the acceptable fits.\footnote{The exact range depends on the assumptions
we make for the acceptable range around each set of ``best-fit'' parameters
Figure~15 of \citetalias{p9-hypothesis} gives, and the actual parameter search we
performed was based the slightly different range $1.50' < v_\text{ref} < 2.26'$.
} This means that only a hollow cone in our $r,v_x,v_y$ parameter space needs to be explored.

While in theory there is a continuum of possible parameter values inside this cone, in
practice the limited angular resolution of the telescope means that very similar parameters
are indistinguishable. From equation~\ref{eq:parallax1} we see that getting the distance
wrong by $\delta r$ results in a parallax ellipse that's bigger by
\begin{align}
	\delta \theta_r &\approx -0.14' \cdot \frac{\delta r}{10 \text{AU}} \Big(\frac{500 \text{AU}}{r}\Big)^2 \label{eq:par-rad}.
\end{align}
Thus shift-stacking with the wrong distance leaves a residual ellipse with a radius of $|\delta\theta_r|$.
If we step through distances in steps of $\Delta r$, then $\delta r$ will take on values in the range
$[-\frac{\Delta r}{2},\frac{\Delta r}{2}]$. Using equations~\ref{eq:par-rad} and \ref{eq:smear-circ} from
Appendix~\ref{sec:smearing}, we see that
on average, this increases the beam FWHM \emph{in quadrature} by:
\begin{align}
	\Delta\text{FWHM}_\text{r} &= 0.093' \cdot \frac{\Delta r}{10 \text{AU}} \Big(\frac{500 \text{AU}}{r}\Big)^2.
\end{align}
Similarly, getting the speed wrong by $\delta v$ will over a time $T = t-t_0$
accumulate to a position error of
\begin{align}
	\delta \theta_v &= 0.3' \cdot \frac{\delta v}{0.1'\text{/yr}} \frac{T}{3\text{yr}}.
\end{align}
For a velocity step of $\Delta v$ we get, using eq.~\ref{eq:smear-lin},
\begin{align}
	\Delta \text{FWHM}_v &= 0.29' \cdot \frac{\Delta v}{0.1'\text{/yr}} \frac{T}{3\text{yr}}
\end{align}
where we have included a factor $\sqrt2$ in the numerical factor to take into account
the smearing in both the $x$ and $y$ directions. The factor $T$ depends on when in the
ACT observing campaign each observation was taken, but will at most be three years if we
choose $t_0$ to be the mid-point of ACT observations.
The integration time $\Delta t$ also results in smearing,
\begin{align}
	\Delta \text{FWHM}_t &= 0.080' \cdot \frac{500\text{AU}}{r} \frac{\Delta t}{\text{day}}
\end{align}
as does the pixel window
\begin{align}
	\Delta \text{FWHM}_\text{pix} &= 0.68' \cdot \frac{\text{res}}{1'}
\end{align}
where {\bf res} is the pixel side length.\footnote{
	This includes a factor $\sqrt2$ because the pixels smear in 2 dimensions,
	but also a factor $1/\sqrt{2}$ because the noise also is being smoothed,
	counteracting some of the S/N loss. This factor is only exactly $1/\sqrt2$
	when smoothing white noise with a Gaussian beam, but numerical tests show that is an excellent approximation even for the top-hat smoothing effect of
	pixel binning.
}\footnote{
	In principle there is also some S/N loss associated with the linear interpolation
	we use during shifting, but this is overwhelmed by the other effects.
} Together these effects make up our \emph{smearing budget}, and each must be chosen small
enough that their combined effect does not overly degrade the S/N. We choose
\begin{itemize}
	\item $\Delta v = 0.1'/\text{yr } \Rightarrow \Delta \text{FWHM}_v = 0.29'$ for $T = 3\text{ yr}$.
	\item $\Delta t = 3 \text{ days } \Rightarrow \Delta \text{FWHM}_t = 0.40'$ for $r = 300 \text{AU}$,
		which is the closest distance we will consider.
	\item $\Delta r = 33 \text{AU} \cdot \left(\frac{500 \text{AU}}{r}\right) \Rightarrow
		\Delta \text{FWHM}_r = 0.31'$. This results in the discrete set of distances
		300, 321, 346, 375, 409, 450, 500, 563, 643, 750, 900, 1125, 1500 and 2000 AU.
		The last two distance bins are more distant than Planet~9 is likely to be, but are
		included because of their low computational cost.
	\item $\text{res} = 1' \Rightarrow \Delta \text{FWHM}_\text{pix} = 0.68'$. That is,
		we use a pixel size of 1 arcmin.\footnote{
			In practice raw maps were built at 0.5$'$ resolution, and were only downsampled (by averaging
			blocks of $2\times2$ pixels) to 1$'$ resolution in the matched filter (that is,
			the $\rho$ and $\kappa$ maps were downsampled). Working with higher resolution
			until this point reduces the aliasing one would otherwise get from working with pixels
			of comparable size to the FWHM.
		}
\end{itemize}
These combine in quadrature to $\Delta \text{FWHM} = 0.90'$, which when combined with our beams represents a
9/19/34\% increase in beam size and loss in S/N in the f090/f150/f220 bands respectively.
The largest contribution to this is the 1 arcmin pixel size. With a $0.5'$ pixel size these
numbers would instead have been 5/11/21\% at a cost of 4$\times$ as high CPU and memory budgets.
We might consider using smaller pixels when we revisit this in the future.

The full, quantized search space is visualized in Figure~\ref{fig:search}. In total the
$r,v_x,v_y$ parameter space has 25\,837 cells.

\subsection{Shift-and-stack implementation}
\label{sec:implementation}
After splitting the 2013--2019 ACT data set into 3-day chunks and building matched filter
maps for each, we were left with 3834 pairs of $\rho$ and $\kappa$ maps taking up a total
of 1.9 TB of disk space. Since these maps are in units of equivalent flux density at the reference
frequency $\nu_0 = 150$ GHz, maps from different arrays and bandpasses that were observed
at the same time, and hence all have the same shifts, can be directly combined before the
main shift-and-stack search. This resulted in a more manageable 787 pairs taking up 220 GB.

The analysis was performed in $10^\circ\times 10^\circ$ tiles with an additional $1^\circ$
padding on all sides using data ``belonging'' to neighboring tiles to avoid discontinuities at tile edges.
For each tile we loop over our parameter space and keep track of the highest-S/N value of
$v_x$ and $v_y$ in each pixel for each value of $r$. This is illustrated in the pseudo-code below:
\begin{lstlisting}[mathescape=true]
for each tile in tiles:
	results = []
	for each r in rs:
		initialize result
		for each $v_x$, $v_y$ given r:
			initialize $\rho_\text{tot}$, $\kappa_\text{tot}$ maps to zero
			for each T, $\rho$, $\kappa$ in tile:
				$\rho_\text{tot}$ += shift($\rho$,r,$v_x$,$v_y$,T)
				$\kappa_\text{tot}$ += shift($\kappa$,r,$v_x$,$v_y$,T)
			update(result,$\rho_\text{tot}$,$\kappa_\text{tot}$,$v_x$,$v_y$)
		results.append(result)
\end{lstlisting}
Here $r$ takes on the values 300, 321, 346, 375, 409, 450, 500, 563, 643, 750, 900,
1125, 1500 and 2000 AU. For each value we visit all velocities $v_x = i\Delta v, v_y = j\Delta v$ where
$i$ and $j$ are integers and
\begin{align}
	1.6'/\text{yr} \cdot \left(\frac{r}{500 \text{AU}}\right)^{-2}\hspace*{-3mm} - \Delta v \le \sqrt{v_x^2+v_y^2} \le 2.3'/\text{yr} \cdot \left(\frac{r}{500 \text{AU}}\right)^{-2} \notag
\end{align}

The function \texttt{shift} applies the coordinate transformation from observed coordinates
at time $t = t_0+T$ to heliocentric coordinates at time $t_0$, taking into account
both parallax for the distance $r$ and the planet's angular velocity $v_x,v_y$.
We use bilinear interpolation to allow for fractional pixel shifts.
This function is the most time-critical part of the search, so it was
implemented in optimized C using AVX intrinsics and OpenMP parallelization.
Since the distance and direction each pixel
is displaced changes slowly as a function of position in the map, we use the same
displacement for blocks of $8\times8$ pixels, saving a large number of trigonometric
operations at no loss of S/N. Overall our implementation is 480 times faster than a straightforward
\texttt{numpy}/\texttt{scipy} implementation.

The function \texttt{update} updates \texttt{result} to maintain a running record of the
highest S/N observed in each pixel, and what value of $\rho_\text{tot}$, $\kappa_\text{tot}$,
$v_x$ and $v_y$ that occured for. We maintain one such \texttt{result} for each value
of $r$ because both bias from mean sky subtraction and the appropriate S/N threshold for
a detection (which depends on the effective number of trials) depend on $r$.

\subsection{Simulations and debiasing}
\label{sec:bias}
Mean sky subtraction mainly removes the static parts of the sky, but it also
subtracts some of the signal from moving objects. These appear as a smeared-out
tracks in the mean sky map, and since part of an object's track necessarily overlaps with
its position in each individual exposure, mean sky subtraction will always lead to
a loss of signal power. The size of the bias is both distance-dependent (because more distant objects
move less and hence overlap more with the mean sky) and position-dependent (because
areas with less coverage will see less of the object's motion).

To map this out we considered
a set of fake planets in a 0.5$^\circ$ grid in heliocentric RA, dec at $t=t_0$, all
with the same flux density but with $r$ stepping through the 14 values we consider in the
parameter search for every 14 grid positions in RA, and $v$ taking on the corresponding
14 values 1.80, 1.57, 1.35, 1.15 , 0.97, 0.80, 0.65 , 0.51, 0.39, 0.29 , 0.20, 0.13, 0.07 and 0.04 $'$/yr.
The direction of the velocity was constant per row, but rotated by 45$^\circ$ for each row.
The result is that all distances are represented in each $\Delta\alpha = 7^\circ,
\Delta\delta = 0.5^\circ$ block of the sky, and all distance-direction combinations are
represented in each $\Delta\alpha = 7^\circ, \Delta\delta = 4^\circ$ block on the sky.

These were used to build new $\rho$ maps $\rho_\text{sim}^\text{raw} = R^T w \circ m_\text{sim}$, where
$\circ$ is the element-wise product, $m_\text{sim}$ is a noise-free map with the simulated sources
in $\micro$K at their observed positions, and $w$ is the white noise inverse variance map
from Section~\ref{sec:skysub}.\footnote{Indices for the individual time-chunks and bandpasses have
been suppressed here for readability.}
Using $w \circ m_\text{sim}$ instead of eq.~\ref{eq:rho}
is an approximation, but based on a small number of full time-domain simulations it appears to be
accurate to $<5\%$. To capture the effect of mean sky subtraction we define the mean flux density map
\begin{align}
	F_\text{mean} &= \frac{\sum_i \rho_{\text{sim},i}}{\sum_i \kappa_i},
\end{align}
where $i$ loops over all the individual maps and the division is element-wise. This was then used
to define the mean sky subtracted simulations:
\begin{align}
	\rho_\text{sim} &= \rho_\text{sim}^\text{raw} - \kappa F_\text{mean},
\end{align}
This mean sky subtraction was done individually for each bandpass, both to avoid mixing
maps with different beams and to reflect what was done to the actual data.

\begin{figure}[htbp]
	\centering
	300 AU \\
	\includegraphics[width=\columnwidth]{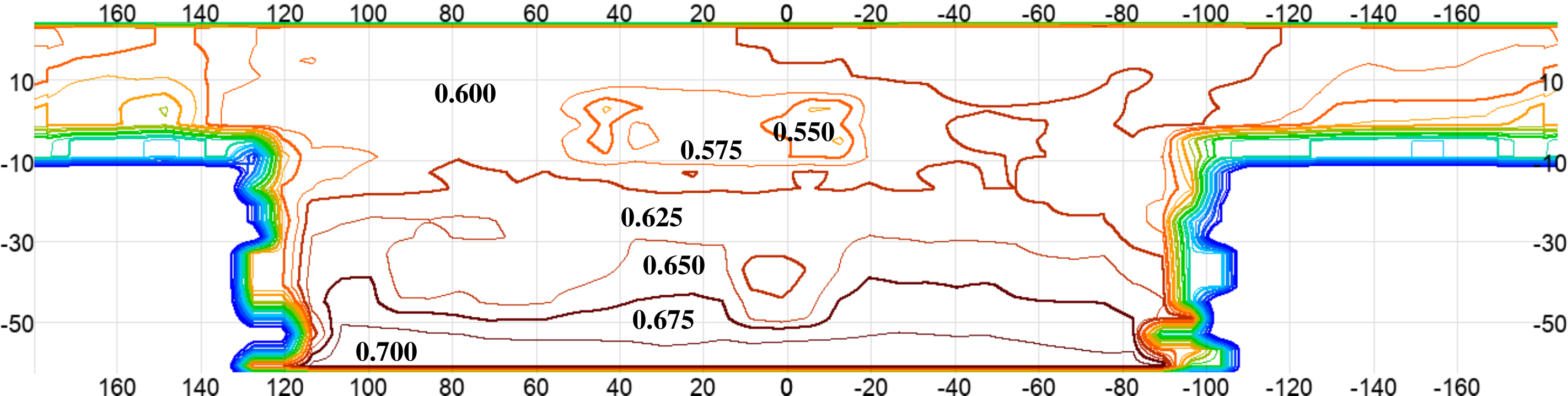} \\
	2000 AU \\
	\includegraphics[width=\columnwidth]{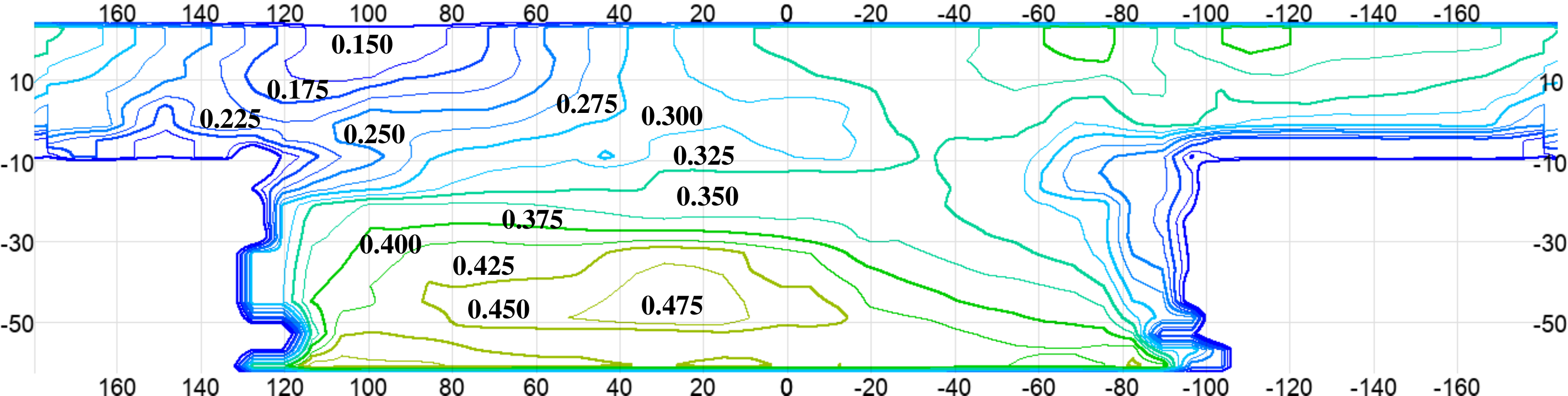}
	\caption{Contour plot of the bias factor recovered from the simulations
	described in Section~\ref{sec:bias}. This is defined as the fraction of
	the true flux density that is recovered. The source of the bias is the
	mean sky subtraction described in Section~\ref{sec:skysub}. The top panel shows the bias for sources
	at 300 AU; the bottom at 2000 AU. The color scale goes from 0 (blue; all
	flux density lost) to 0.70 (red, 70\% of flux density recovered), with a contour interval
	of 0.025. The horizontal and vertical axes are RA and dec respectively.
	We divide by these factors to debias the recovered flux densities.
	}
	\label{fig:bias}
\end{figure}

Finally, we ran the shift-and-stack procedure from Section~\ref{sec:implementation} on
the simulated data set, and read off the recovered flux density for each simulated source.
We find practically no dependence on the direction of the velocity, and therefore
average the data points for different velocities for our final bias model, resulting
in bias maps $\vec b(r)$ with resolution $7^\circ \times 4^\circ$. These are shown for the
closest and furthest Planet~9 distance considered in Figure~\ref{fig:bias}. The bias changes
smoothly with position and is well resolved even with these large pixels. The standard
deviation of the data points going into each pixel is about 0.5\%, which we take as the
uncertainty on our bias maps. We use this to define $\rho_\text{tot}^\text{debiased}
= \rho_\text{tot} \circ \vec b$ and $\kappa_\text{tot}^\text{debiased} = 
\kappa_\text{tot} \circ \vec b^2$, from which bias-free flux densities can be recovered
via eq.~\ref{eq:avec}.

\subsection{Significance}
\label{sec:norm}
Our search method results in a map for each $r$ where each pixel has the maximum
S/N across all the velocity parameters for that $r$. To construct a list of
detection candidates and detection limit maps we need to know the background
distribution of these S/N values. This is made difficult by the varying depth, and varying temporal and
spatial distribution of the data used in the search. The effective number of trials is a
strong function of $r$, and the individual trials are correlated, with
the correlation depending on how densely the ACT observations covers each
spot of the map. The S/N distribution should therefore vary both as a function
of $r$ and position.

The simple approach of multiplying the number of beams
in the map ($\sim 30$ million) with the total trial number (25\,837)
to get a total number of trials ($\sim 10^{12}$) and a corresponding
Gaussian quantile ($7 \sigma$) does not work. Aside from
overestimating the effective number of trials, it would also lead to
the search grossly preferring candidates with low $r$ by not penalizing
the much larger parameter space for low $r$ compared to high $r$.

Instead we will take the approach of transforming S/N into an overall
detection statistic $z$ that follows a simple, uniform Gaussian
distribution, at least for its high-$z$ tail. This procedure is
described in Appendix~\ref{sec:norm-app}, where we find that
\begin{align}
	z &= \xi(S/N) = (S/N-\mu_z)/\sigma_z
\end{align}
where $\mu_z$ and $\sigma_z$ are functions of distance $r$ and position
in the map.

\subsection{Candidate identification}
With the normalized detection statistic $z$ in hand, we build a
set of preliminary candidate detections by selecting peaks with
$z>3.5$. Given the large sky area covered, this low threshold
will result in a large number of candidates, the vast majority of
which would of course simply be noise fluctuations (especially
considering that we expect at most one real object), but that
allows us to get a good handle on the background distribution
that any real objects would stand out from.

To better understand the background, we took advantage that the planet signature
would be positive in our maps and repeated the whole search with the
sign of all the data flipped.
No signal is expected in the sign-flipped search, but it shares the same noise properties
and many of the systematics (e.g., variable point sources and edge artifacts),
so it gives a good estimate of the background detection rate.

\begin{figure}[htbp]
	\centering
	\hspace*{-4mm}\includegraphics[width=1.15\columnwidth]{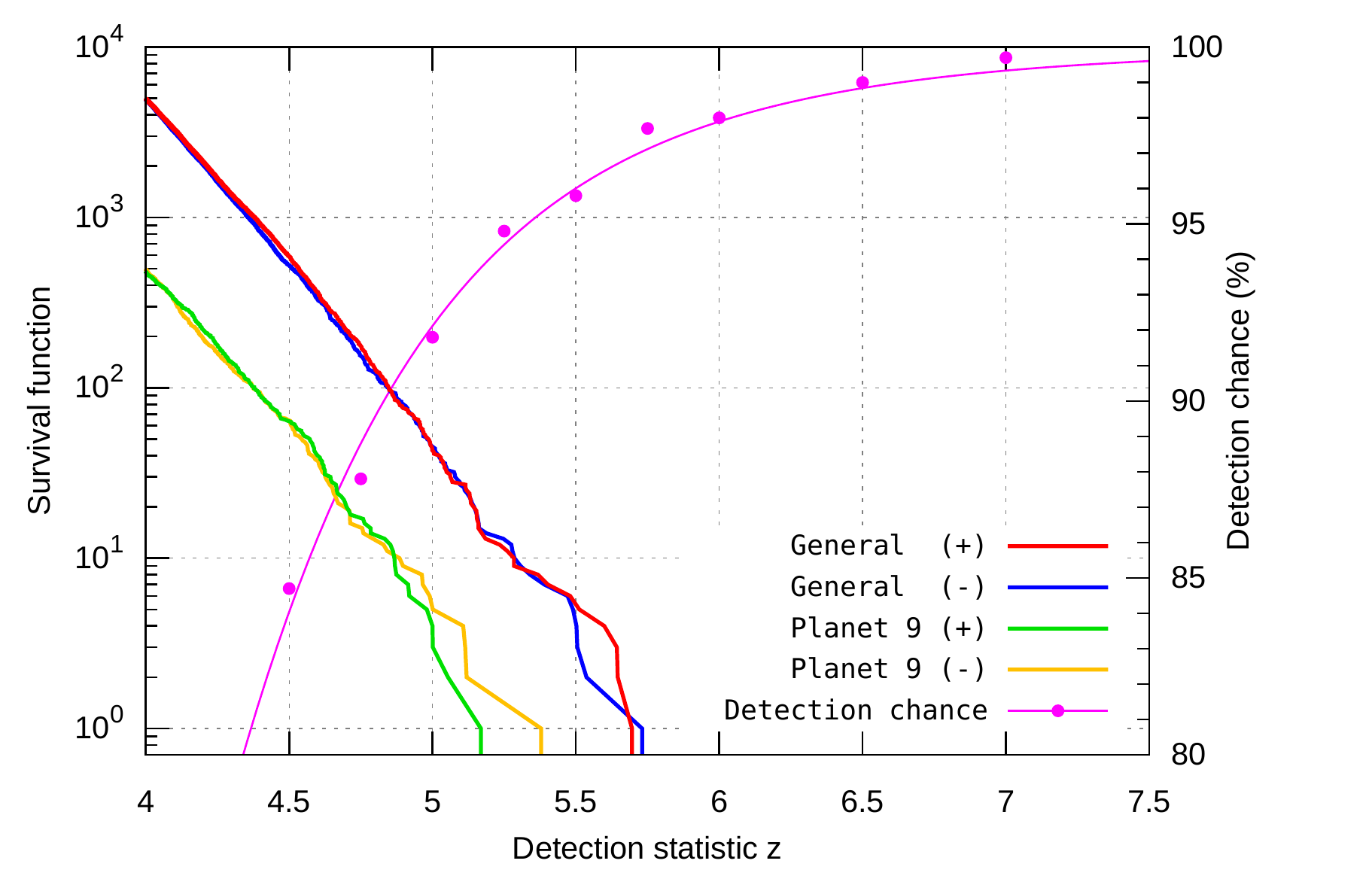}
	\caption{\dfn{Left axis}: Distribution of candidate detections for the Planet~9-like (green)
	and general (red) search compared to sign-flipped versions of the
	same searches (blue and yellow respectively), as a function of
	the detection statistic $z$. No signal is expected
	in the sign-flipped search, but it shares the same noise properties and
	many of the systematics (e.g., variable point sources and edge artifacts),
	so it gives a good estimate of the background detection rate.
	The lack of excess events in the positive curves vs. the negative ones
	(beyond the scatter expected from Poisson sample variance) means that
	we do not have any significant detections.
	\dfn{Right axis}: The probability of recovering an injected object as
	a function of $z$ (magenta). The detection probability is 95\% by $z=5.3$.
	}
	\label{fig:cand-sf}
\end{figure}

We classified each candidate as \emph{Planet~9-like} or \emph{general}
based on whether they satisfied the expected bounds on Planet~9's
orbital inclination, $10^\circ < i < 30^\circ$ \citepalias{p9-hypothesis}.\footnote{Note
that this inclination bound is the only difference between the ``Planet~9-like''
and ``general'' categories. Because both of them are based on a parameter search
that only considered distances and velocities reasonable for Planet~9 (see Section~\ref{sec:search-space}),
even the ``general'' search is not sensitive to planets with extreme ellipticity or
$r < 300 \text{ AU}$.}
The inclination is not one of the free parameters of our fit,
but we can approximate this selection by transforming the candidate
coordinates and velocities into ecliptic coordinates, and requiring
\begin{align}
	\quad 10^\circ &< \hat i < 30^\circ &&,&
	\hat i &= \sqrt{\beta^2 + v_\beta^2/v_\lambda^2}
\end{align}
and where $\lambda,\beta$ are the ecliptic longitude and latitude respectively, and
$v_\lambda,v_\beta$ are the velocity components in those directions.\footnote{
	In practice we accidentally used $v_\beta' = \text{max}(|v_\beta|-\Delta v,0)$
	and $v_\lambda' = \text{max}(|v_\lambda|-\Delta v,0)$ instead. These were supposed
	to avoid division by zero, but by using the wrong sign in front of $\Delta v$
	they instead increased the likelihood for this. In practice this has negligible effects
	on our results, since only the highest distance bin $r = 2000$ AU has low enough
	speeds that a 0.05$'$/yr difference would matter.
} The formula for $\hat i$ assumes that orbits have $\beta(\lambda) = i\sin(\lambda-\lambda_0)$,
which is a decent approximation as long as $i$ is small.

Finally, the top 100 from each list were visually inspected using both the $z$ maps,
the best-fit shift-stacked maps, raw sky maps and individual 3-day matched filter maps,
and any obvious problems like
edge artifacts, uncut variable point sources etc. were cut.\footnote{Below the top 100
the statistics are completely dominated by noise fluctuations, and any artifacts would
be hard to distinguish from noise anyway because of the low S/N.}

\subsection{Flux and distance limits}
It is useful to be able to translate the survey depth into detection limit maps.
To do this, we need the \emph{false negative rate} as a function of the detection
statistic $z$. We found this by repeating the signal injection, search and detection
procedure from Section~\ref{sec:bias} with two two important differences:
\begin{enumerate}
	\item Simulated sources were \emph{added} to the data instead of replacing it, resulting
		in noisy simulations.
	\item For each simulated source we chose a target
		$z \in $ \{4.50, 4.75, 5.00, 5.25, 5.50, 5.75, 6.00, 6.50, 7.00, 8.00, 10.00, 15.00\},
		translated this into a S/N ratio using $\xi^{-1}(z)$ (see Section~\ref{sec:norm}
		and Appendix~\ref{sec:norm-app}), and combined it with the local survey depth to
		define a simulated flux density $s = \xi^{-1}(z) / \sqrt{\kappa_\text{tot}^\text{debiased}}$.
\end{enumerate}
We then ran the standard mean sky subtraction and candiate search on the maps,
and computed the fraction of the injected sources that were ultimately recovered
as a function of $z$. The result is plotted as the curve ``detection chance''
in Figure~\ref{fig:cand-sf}. Overall we find that a source bright enough to
correspond to $z=5.3$ has a 95\% chance of being detected. Hence, the
95\% flux density detection limit map is given by
\begin{align}
	s_\text{lim}^{95\%} &= \xi^{-1}(5.3) / \sqrt{\kappa_\text{tot}^\text{debiased}}
\end{align}
Aside from its position-dependence this limit is also distance-dependent,
since both $\xi$ and $\kappa_\text{tot}^\text{debiased}$ depend on $r$.

Given a model for Planet~9's luminosity we can translate the flux density limit to a distance
limit. Since the flux density falls with the square of the distance, the distance limit $r_\text{lim}^{95\%}$
can be found as the solution to the equation
\begin{align}
	(r_\text{ref}/r)^2 s_\text{ref} &= s_\text{lim}^{95\%}(r_\text{lim})
\end{align}
with Table~\ref{tab:p9-props} showing examples of the reference flux $s_\text{ref}$ for $r_\text{ref} = 500$ AU for different
Planet~9 scenarios.\footnote{We assume that $s_\text{lim}^{95\%}$ changes linearly between the discrete set
of distances $r \in $\{300, 321, 346, 375, 409, 450, 500, 563, 643, 750, 900, 1125, 1500, 2000\} AU
where we computed it.}

\section{Results}
The search resulted in 38\,000 raw candidates, of which 3\,500 and 35\,000
fell into the Planet~9-like and general categories respectively.
Manual inspection of the top 100 candidates led to 3 Planet~9-like and
17 general candidates being cut. These included the first three transients detected
by ACT, which were published in a separate paper \citep{act-first-transients}.
The top ten candidates from the Planet~9-like and general searches are shown in
Tables~\ref{tab:cands-p9} and \ref{tab:cands-ok}. The full candidate
distribution is shown in Figure~\ref{fig:cand-sf} and is identical to within sample
variance for both the normal and sign-inverted searches.
The lack of excess events in the distribution of normal candidates vs. sign-inverted
candidates means that {\bf we have no statistically significant detections}.

\begin{figure*}[htbp]
	\centering
	$300<r<600$ AU\\
	\includegraphics[width=\textwidth]{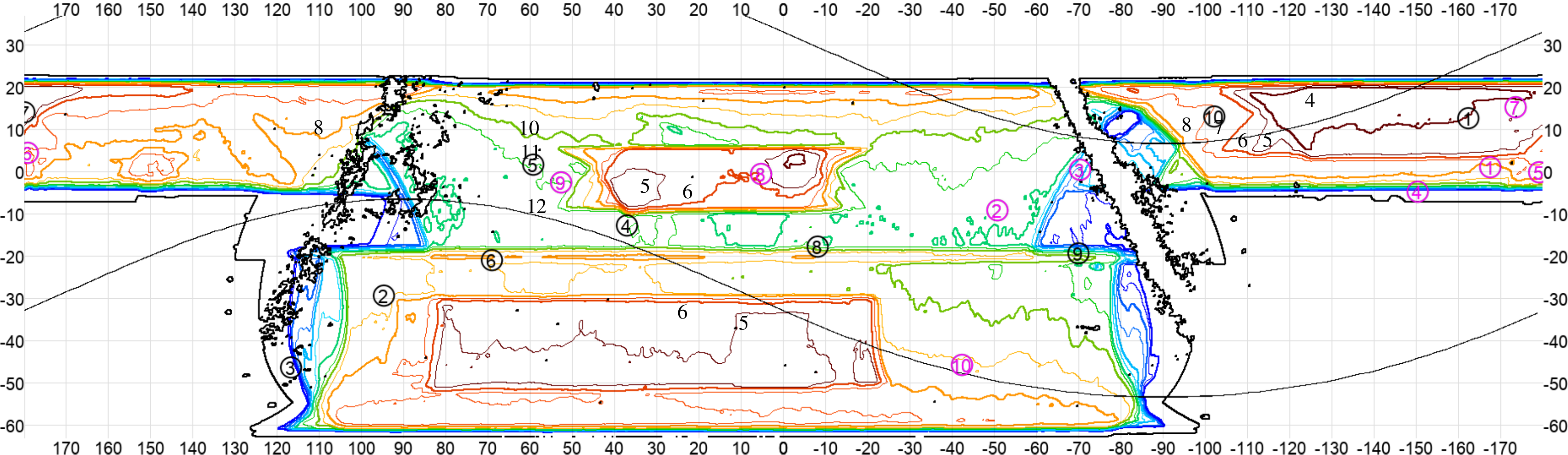}
	$r = 1500$ AU\\
	\includegraphics[width=\textwidth]{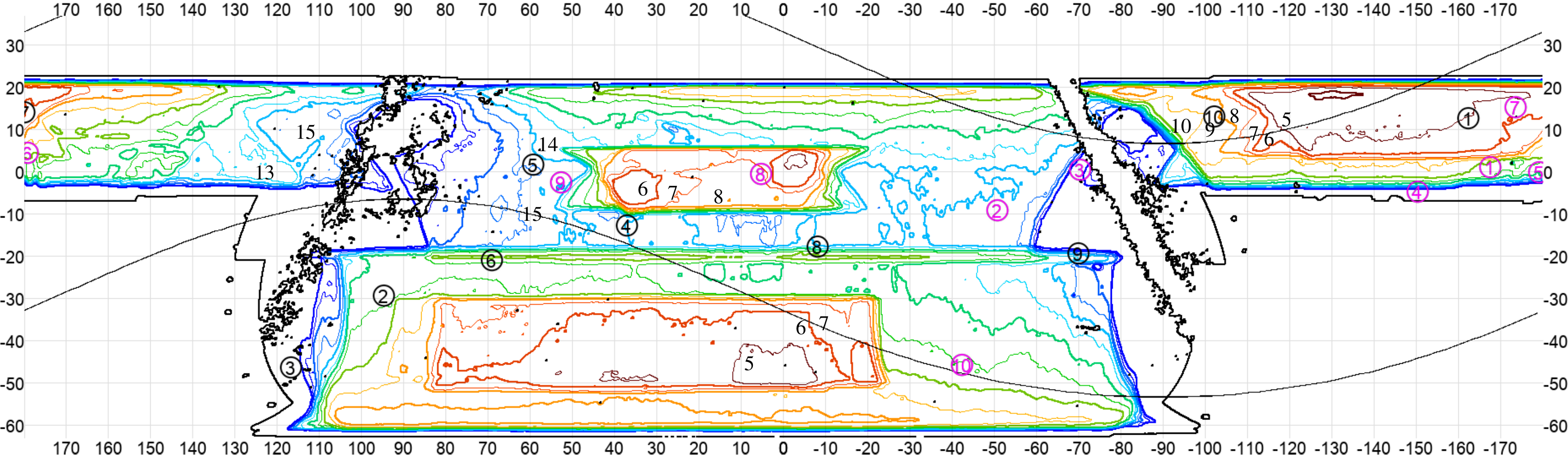}
	\caption{Flux limit for the general Planet~9 search for distance $300<r<600$ AU
	(top) and $r = 2000$ AU (bottom) in equatorial coordinates. Objects brighter than this would be part of
	our top-10 candidate list. The contours go from red (4 mJy) to
	blue (20 mJy) in steps of 1 mJy. Some individual contour lines are labeled
	(plain black numbers)
	with their corresponding depth, for convenience. The flux density limit depends on the size of
	the parameter search space, which shrinks with $r$ leading to a lower
	flux density limit; and the loss from mean sky subtraction, which raises the
	flux density limit. These effects mostly balance each other for $300<r<600$,
	while the mean sky subtraction loss dominates at higher $r$, leading to
	a gradually increasing flux density limit. The circled numbers show the locations
	of the top-10 candidates from the general (black) and Planet~9-like (magenta)
	searches. The thin black curves delimit the area with inclination less than 30$^\circ$.
	The point source mask was left out from this plot, but its
	effect can be seen in Figure~\ref{fig:rlim}.
	}
	\label{fig:fluxlim}
\end{figure*}

Given our non-detection, we constrain the flux density from Planet~9 or similar objects in the
outer solar system to be <4--12 mJy (95\% confidence) for $r \ge 300\text{ AU}$
inside our survey area, depending on local survey depth. This limit is approximately
distance-independent in the range $300\text{ AU} \le r \le 600\text{ AU}$, after which
it gradually worsens to <5--15 mJy by 1500 AU. We show a map of the flux density limit in
Figure~\ref{fig:fluxlim}, along with the locations of the top 10 candidates from the
Planet~9-like and general searches.

\begin{figure*}[htbp]
	\centering
	$5 M_\Earth$ (250--625 AU)\\
	\includegraphics[width=\textwidth]{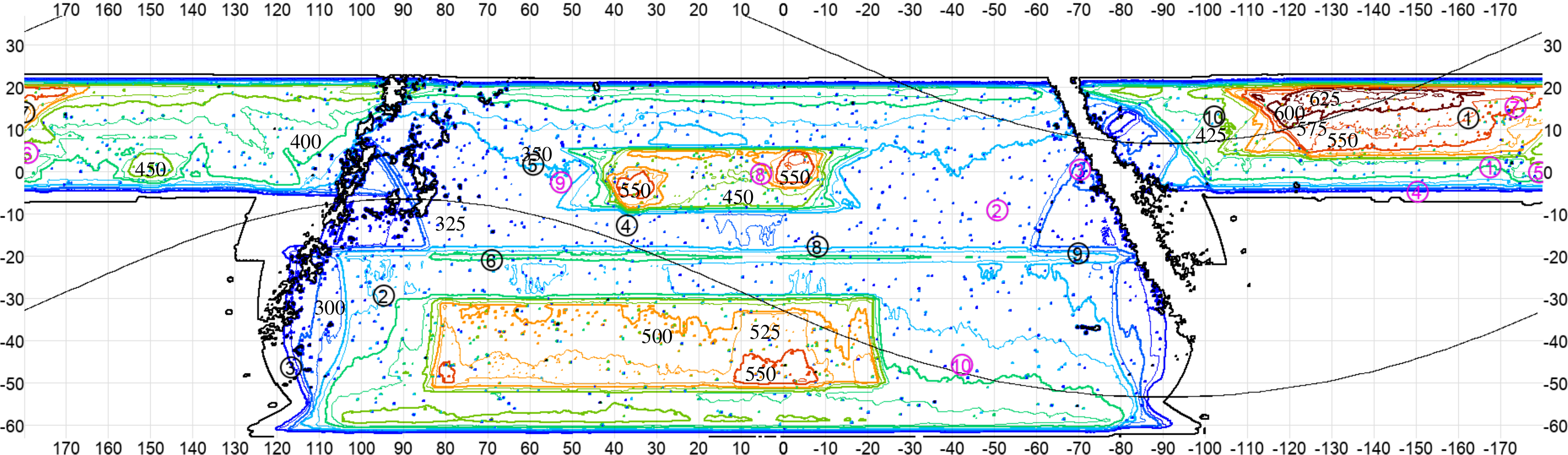}
	$10 M_\Earth$ (350--775 AU)\\
	\includegraphics[width=\textwidth]{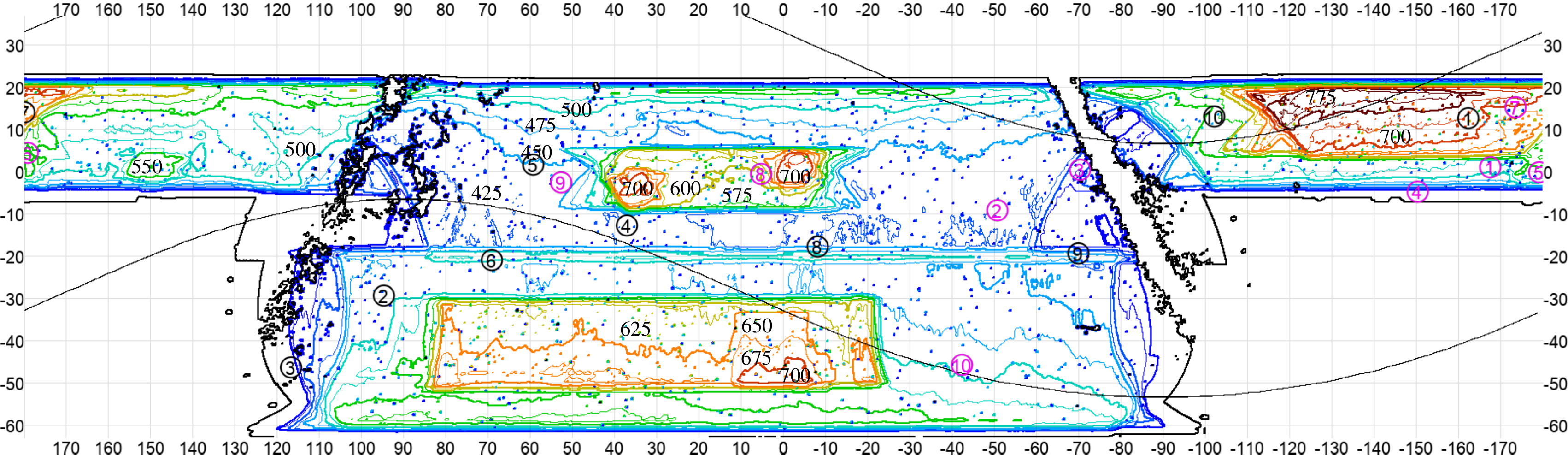}
	\caption{\dfn{Top}: Distance limit for detection of a $5 M_\Earth$
	Planet~9 with the nominal composition from Section~\ref{sec:p9-phys}. Contours go from 250 AU
	(blue) to 550 AU (red) in steps of 25 AU, with some contours
	labeled for convenience. Note that $r<300$ AU were not
	included in the search, so areas with a distance limit $\lesssim
	300$ AU do not meaningfully constrain Planet~9.
	The circled numbers show the locations
	of the top-10 candidates from the general (black) and Planet~9-like (magenta)
	searches. The little colored dots are caused by the point source
	mask. Its effect appears exaggerated because of the low resolution of
	the plot -- in reality this mask only affects a tiny fraction of the sky. \dfn{Bottom}:
	As above, but for a $10 M_\Earth$ Planet~9. The contours here go from
	350 AU (blue) to 750 AU (red).
	}
	\label{fig:rlim}
\end{figure*}

Figure~\ref{fig:rlim} shows the corresponding distance limits for the nominal
$5 M_\Earth$ and $10 M_\Earth$ scenarios from Section~\ref{sec:p9-phys}. In the
shallower parts of our survey area, a $5 M_\Earth$ Planet~9 would need to be
more distant than 325 AU to evade detection. This increases to 625 AU in the
deepest parts of our survey. For a $10 M_\Earth$ planet these numbers increase to
425 AU and 775 AU respectively.

We cover quite low galactic latitudes, but parts of the galaxy is still masked. This is usually
confined to $|b|<2.5^\circ$, but it is not uncommon for the mask to extend beyond this to cover
features like the Orion Nebula.

\begin{table*}[p]
	\centering
	\begin{closetabrows}[0.5]
	\begin{tabular}{r>{\centering\arraybackslash}m{15mm}@{\hspace{0.5mm}}>{\centering\arraybackslash}m{15mm}@{\hspace{0.5mm}}>{\centering\arraybackslash}m{15mm}@{\hspace{0.5mm}}>{\centering\arraybackslash}m{15mm}rrrrrrrr}
		\bf \# & \bf z map & \bf Stack & \bf f090 & \bf f150 & \bf RA & \bf Dec & $\pmb{z}$ & \bf F & $\pmb{\Delta}$\bf{}F & $\pmb{r}$ & $\pmb{v_x}$ & $\pmb{v_y}$ \\
		& & & & & \scriptsize ($^\circ$) & \scriptsize ($^\circ$) & & \scriptsize (mJy) & \scriptsize (mJy) & \scriptsize (AU) & \scriptsize ($'$/yr) & \scriptsize ($'$/yr) \\
		\noalign{\vspace*{0.7mm}}
		\hline
		\noalign{\vspace*{0.7mm}}
  1 & \includegraphics[height=15mm]{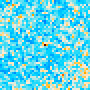} & \includegraphics[height=15mm]{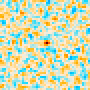} & \includegraphics[height=15mm]{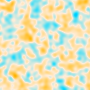} & \includegraphics[height=15mm]{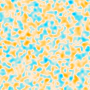} & -167.54 &    1.04 &    5.17 &    8.3 &    1.8 &  375 &   2.2 &  -2.9 \\
  2 & \includegraphics[height=15mm]{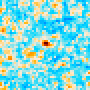} & \includegraphics[height=15mm]{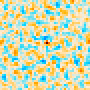} & \includegraphics[height=15mm]{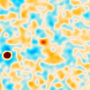} & \includegraphics[height=15mm]{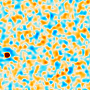} &  -50.84 &   -9.16 &    5.05 &   11.5 &    2.4 &  375 &   0.2 &   3.0 \\
  3 & \includegraphics[height=15mm]{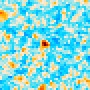} & \includegraphics[height=15mm]{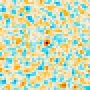} & \includegraphics[height=15mm]{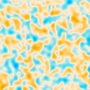} & \includegraphics[height=15mm]{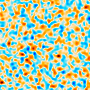} &  -70.32 &    0.34 &    5.00 &   14.8 &    3.1 &  321 &   0.6 &   4.5 \\
  4 & \includegraphics[height=15mm]{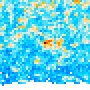} & \includegraphics[height=15mm]{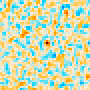} & \includegraphics[height=15mm]{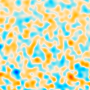} & \includegraphics[height=15mm]{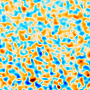} & -150.37 &   -4.86 &    5.00 &   23.2 &    5.5 &  643 &  -0.1 &  -1.1 \\
  5 & \includegraphics[height=15mm]{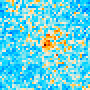} & \includegraphics[height=15mm]{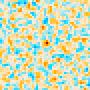} & \includegraphics[height=15mm]{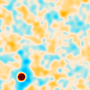} & \includegraphics[height=15mm]{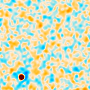} & -179.17 &   -0.23 &    4.98 &    8.5 &    1.9 & 1125 &   0.0 &   0.4 \\
  6 & \includegraphics[height=15mm]{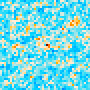} & \includegraphics[height=15mm]{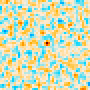} & \includegraphics[height=15mm]{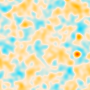} & \includegraphics[height=15mm]{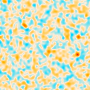} &  179.01 &    4.34 &    4.92 &    6.3 &    1.4 &  500 &   1.3 &  -1.8 \\
  7 & \includegraphics[height=15mm]{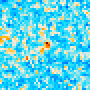} & \includegraphics[height=15mm]{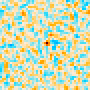} & \includegraphics[height=15mm]{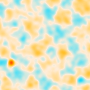} & \includegraphics[height=15mm]{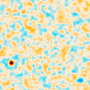} & -173.55 &   15.20 &    4.92 &    4.1 &    0.9 &  346 &  -0.7 &   4.1 \\
  8 & \includegraphics[height=15mm]{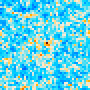} & \includegraphics[height=15mm]{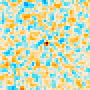} & \includegraphics[height=15mm]{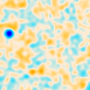} & \includegraphics[height=15mm]{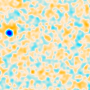} &    5.25 &   -0.70 &    4.87 &    5.6 &    1.3 &  643 &  -0.1 &   1.3 \\
  9 & \includegraphics[height=15mm]{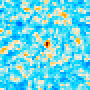} & \includegraphics[height=15mm]{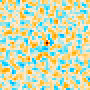} & \includegraphics[height=15mm]{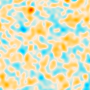} & \includegraphics[height=15mm]{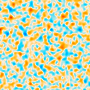} &   52.66 &   -2.60 &    4.87 &   10.1 &    2.3 &  563 &  -0.2 &  -1.7 \\
 10 & \includegraphics[height=15mm]{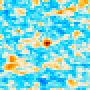} & \includegraphics[height=15mm]{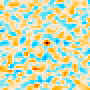} & \includegraphics[height=15mm]{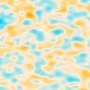} & \includegraphics[height=15mm]{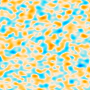} &  -42.35 &  -45.80 &    4.87 &    8.4 &    1.7 &  500 &   0.3 &   1.8 \\
	\hline
	\end{tabular}
	\end{closetabrows}
	\caption{Top 10 Planet9-like candidates, sorted by the detection statistic $z$
	(see Section~\ref{sec:norm} or Appendix~\ref{sec:norm-app} for definition). The columns are:
	\dfn{\#}: The rank in terms of peak $z$ value. \dfn{z map}: A thumbnail of the
	$z$ map centered on the candidate. \dfn{Stack}: The shift-and-stack (i.e. motion-corrected)
	map for the best-fit parameters. \dfn{f090/f150}: Filtered versions of the mean sky model in the f090/f150 band.
	Because these do not include any motion correction, no Planet~9 signal is expected here, but
	they are useful for seeing how ``clean'' each candidate's neighborhood is, e.g. if there
	are any bright point sources, dust clumps or map edges at or near the candidate's location.
	All thumbnails are
	$45'\times45'$ centered on the candidates.
	\dfn{RA}, \dfn{Dec}: Candidate's J2000 heliocentric equatorial coordinates on modified
	Julian day (MJD) 57688. \dfn{z}: The candidate's detection statistic $z$. \dfn{F}, \dfn{$\Delta$F}:
	Flux in the f150 band in mJy, assuming a 40 K blackbody, and its uncertainty.
	\dfn{r}: Distance from the Sun, in AU.
	$\pmb{v_x}$, $\pmb{v_y}$: Intrinsic motion in arcmin per year.}
	\label{tab:cands-p9}
\end{table*}

\begin{table*}[p]
	\centering
	\begin{closetabrows}[0.5]
	\begin{tabular}{r>{\centering\arraybackslash}m{15mm}@{\hspace{0.5mm}}>{\centering\arraybackslash}m{15mm}@{\hspace{0.5mm}}>{\centering\arraybackslash}m{15mm}@{\hspace{0.5mm}}>{\centering\arraybackslash}m{15mm}rrrrrrrr}
		\bf \# & \bf z map & \bf Stack & \bf f090 & \bf f150 & \bf RA & \bf Dec & $\pmb{z}$ & \bf F & $\pmb{\Delta}$\bf{}F & $\pmb{r}$ & $\pmb{v_x}$ & $\pmb{v_y}$ \\
		& & & & & \scriptsize ($^\circ$) & \scriptsize ($^\circ$) & & \scriptsize (mJy) & \scriptsize (mJy) & \scriptsize (AU) & \scriptsize ($'$/yr) & \scriptsize ($'$/yr) \\
		\noalign{\vspace*{0.7mm}}
		\hline
		\noalign{\vspace*{0.7mm}}
  1 & \includegraphics[height=15mm]{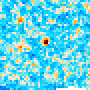} & \includegraphics[height=15mm]{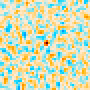} & \includegraphics[height=15mm]{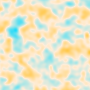} & \includegraphics[height=15mm]{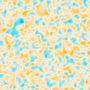} & -162.40 &   12.65 &  5.65 &    4.4 &  0.8 &  300 &   0.7 &   5.9 \\
  2 & \includegraphics[height=15mm]{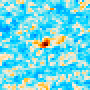} & \includegraphics[height=15mm]{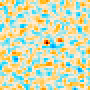} & \includegraphics[height=15mm]{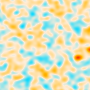} & \includegraphics[height=15mm]{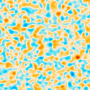} &   94.55 &  -29.48 &  5.64 &    9.7 &  1.8 &  500 &   1.6 &   1.5 \\
  3 & \includegraphics[height=15mm]{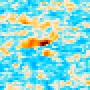} & \includegraphics[height=15mm]{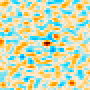} & \includegraphics[height=15mm]{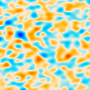} & \includegraphics[height=15mm]{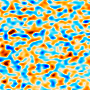} &  116.58 &  -46.50 &  5.60 &   25.1 &  4.3 &  300 &   4.3 &   4.3 \\
  4 & \includegraphics[height=15mm]{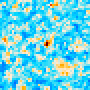} & \includegraphics[height=15mm]{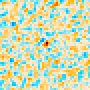} & \includegraphics[height=15mm]{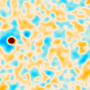} & \includegraphics[height=15mm]{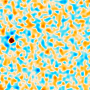} &   36.91 &  -12.81 &  5.51 &   13.7 &  3.4 & 1500 &   0.0 &  -0.1 \\
  5 & \includegraphics[height=15mm]{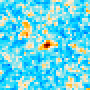} & \includegraphics[height=15mm]{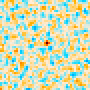} & \includegraphics[height=15mm]{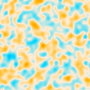} & \includegraphics[height=15mm]{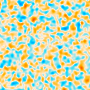} &   59.20 &    1.52 &  5.48 &   11.3 &  2.4 &  643 &   0.6 &   0.6 \\
  6 & \includegraphics[height=15mm]{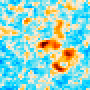} & \includegraphics[height=15mm]{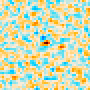} & \includegraphics[height=15mm]{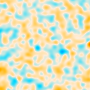} & \includegraphics[height=15mm]{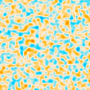} &   69.03 &  -21.10 &  5.40 &    9.0 &  1.7 &  300 &   3.9 &   1.4 \\
  7 & \includegraphics[height=15mm]{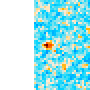} & \includegraphics[height=15mm]{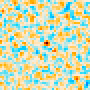} & \includegraphics[height=15mm]{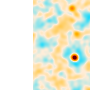} & \includegraphics[height=15mm]{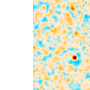} &  179.90 &   13.94 &  5.28 &    4.8 &  0.9 &  346 &  -3.7 &  -2.3 \\
  8 & \includegraphics[height=15mm]{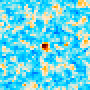} & \includegraphics[height=15mm]{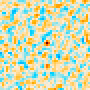} & \includegraphics[height=15mm]{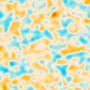} & \includegraphics[height=15mm]{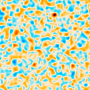} &   -8.19 &  -17.78 &  5.28 &   12.1 &  2.8 & 1125 &  -0.3 &   0.0 \\
  9 & \includegraphics[height=15mm]{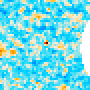} & \includegraphics[height=15mm]{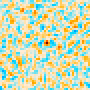} & \includegraphics[height=15mm]{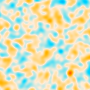} & \includegraphics[height=15mm]{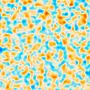} &  -69.90 &  -19.31 &  5.15 &   14.9 &  4.4 & 2000 &   0.0 &   0.1 \\
 10 & \includegraphics[height=15mm]{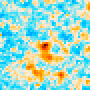} & \includegraphics[height=15mm]{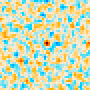} & \includegraphics[height=15mm]{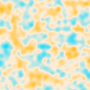} & \includegraphics[height=15mm]{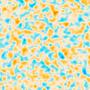} & -102.24 &   13.04 &  5.14 &    6.5 &  1.4 &  500 &  -1.4 &  -1.6 \\
	\hline
	\end{tabular}
	\end{closetabrows}
	\caption{Like Table~\ref{tab:cands-p9}, but for the general candidates.}
	\label{tab:cands-ok}
\end{table*}

\begin{figure*}
	\centering
	\begin{closetabcols}
	\hspace*{-10mm}\begin{tabular}{cc}
	$5 M_\Earth$ & $10 M_\Earth$ \\
	\includegraphics[height=77mm,trim=5mm 0mm 5mm 5mm]{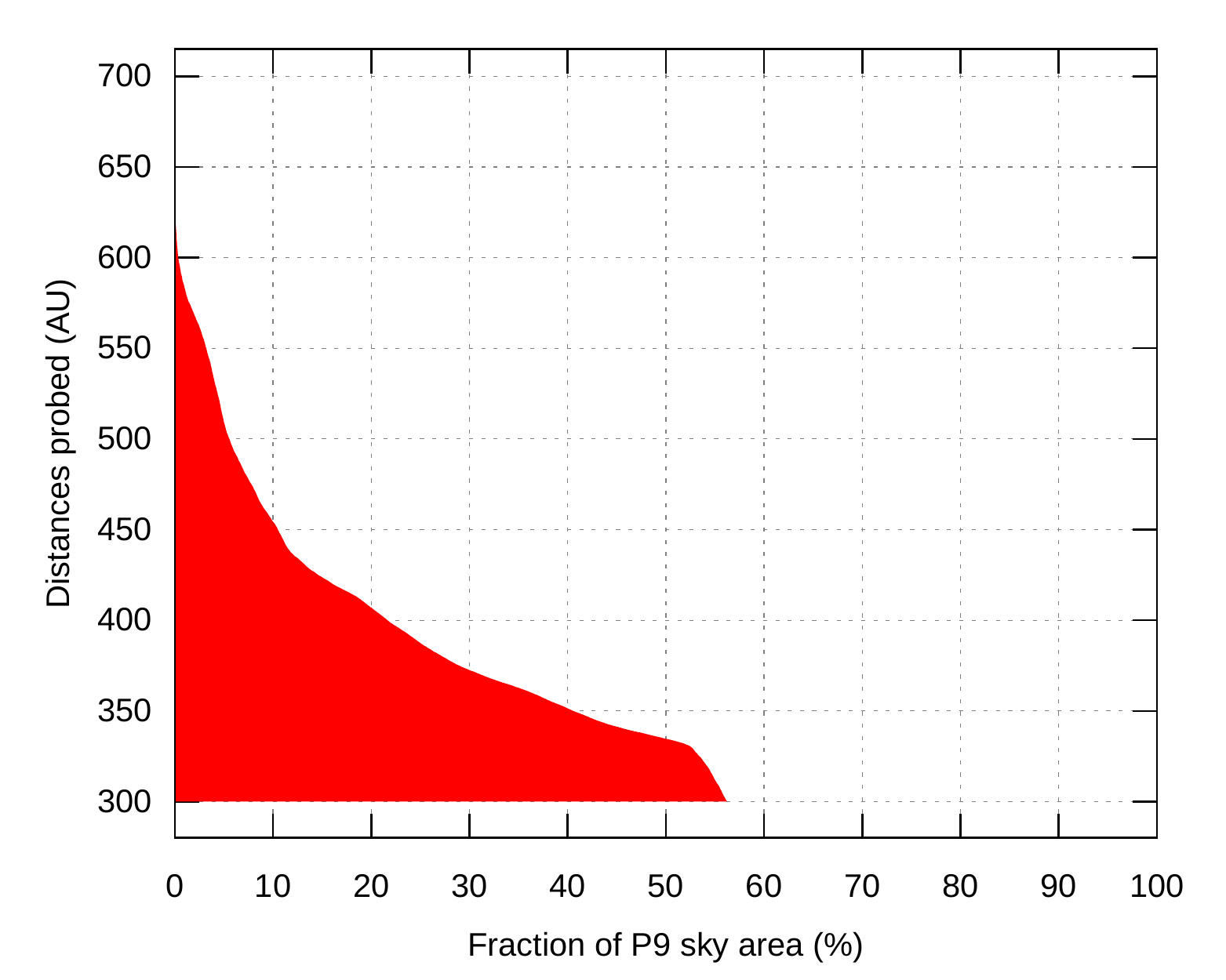} &
	\includegraphics[height=77mm,trim=5mm 0mm 5mm 5mm]{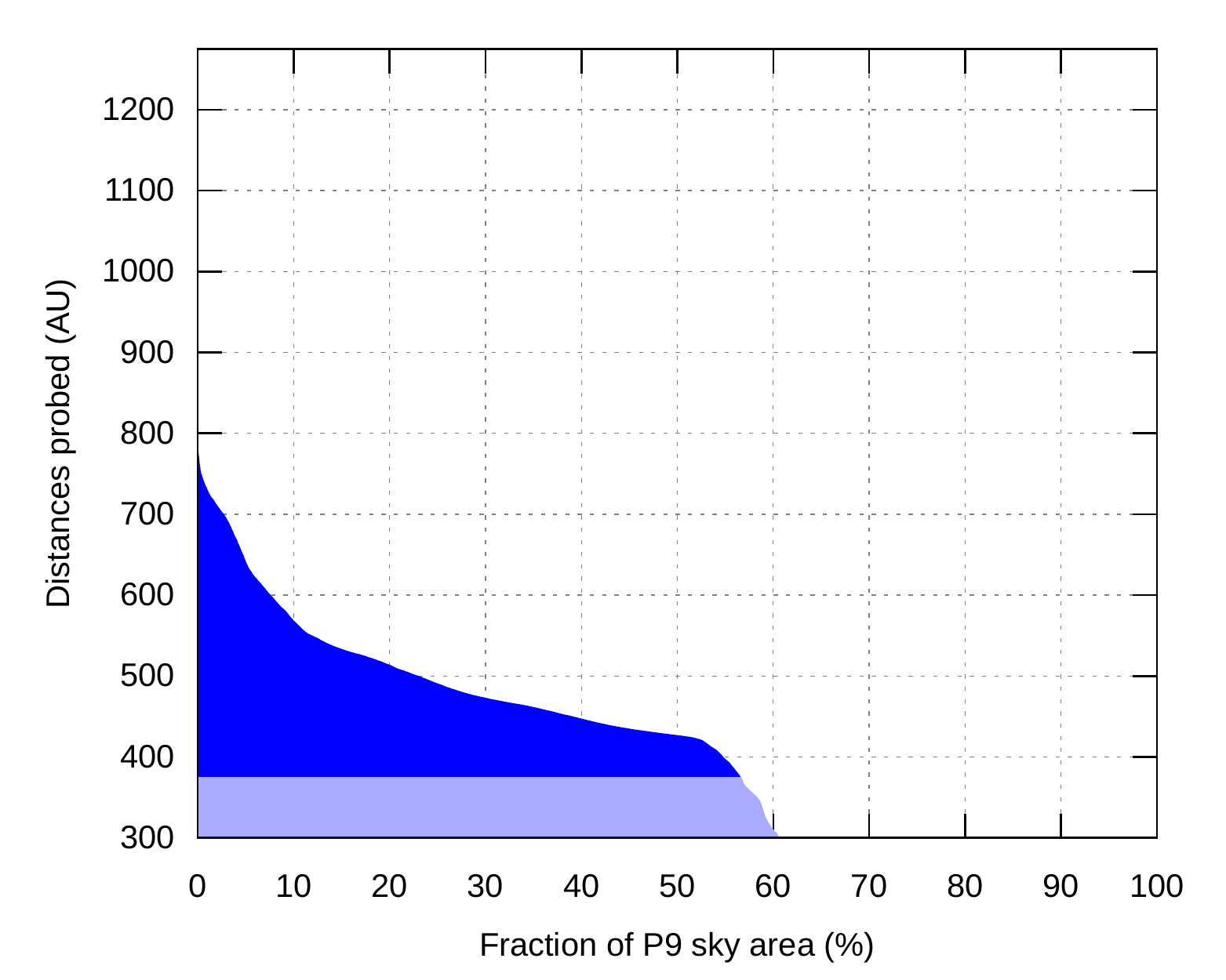}
	\end{tabular}
	\end{closetabcols}
	\caption{Our Planet~9 exclusion limit distribution with the
	nominal composition from section~\ref{sec:p9-phys}, showing what distance
	range we can exclude over a given fraction of the Planet~9 expected sky
	area ($|\beta|<30^\circ$). The value on the x axis gives the
	fraction of the Planet~9 sky area where our bounds are at least as good
	as the distance given on the y axis. This is what one gets if one
	sorts the values in Figure~\ref{fig:rlim} in descending order.
	\dfn{Left}: The nominal $5 M_\Earth$ Planet~9 is predicted to
	be between roughly 280 AU and 715 AU distant (see Table~\ref{tab:prior}).
	This curve shows that we are sensitive up to 600 AU over a few
	percent of the Planet~9 sky area; up to 450 AU over 10\% of
	that area; 300 AU to at least 350 AU over 40\% of the area; and so on.
	The colored fraction of the graph gives a rough idea of what fraction
	of the Planet~9 parameter space we probe.
	\dfn{Right}: As left, but for the $10 M_\Earth$ version of the planet.
	The light shaded region represents distances we probe that are
	closer than what is allowed by the prior.
	}
	\label{fig:rlim-1d}
\end{figure*}

\begin{figure}[ht]
	\centering
	$5 M_\Earth$ \\
	\includegraphics[width=\columnwidth]{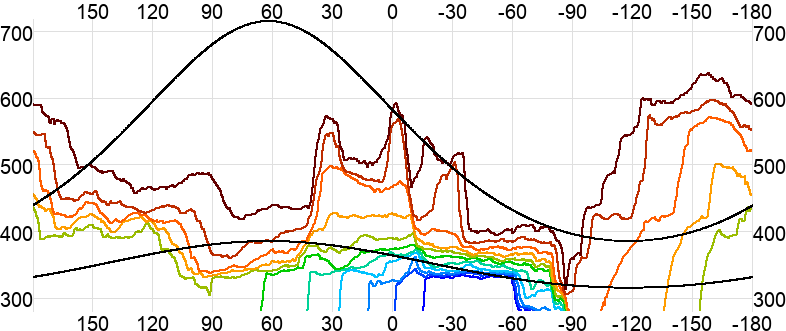} \\
	\vspace*{3mm}
	$10 M_\Earth$ \\
	\includegraphics[width=\columnwidth]{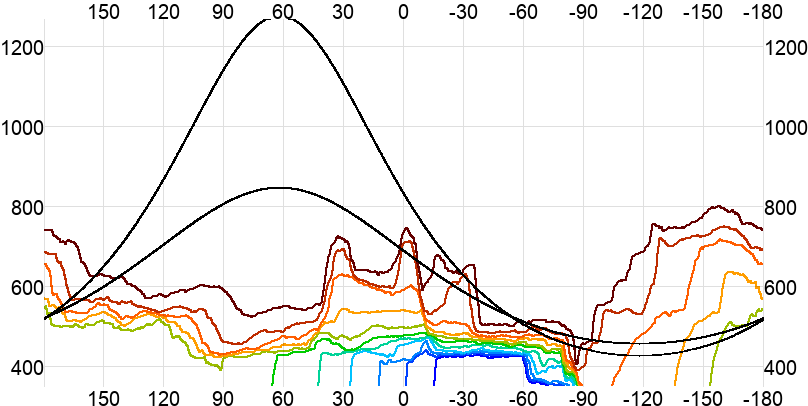}
	\caption{Distance limit distribution as a function of ecliptic longitude
	compared to the Planet~9 orbit. The x axis is ecliptic longitude in degrees.
	The y axis is the planet's current distance from the Sun.
	The colored curves are deciles of the distance limit distribution
	for the Planet~9-like search,
	from blue (0\%) to red (100\%). The bottom and top black curves represent
	respectively a low-$a$, low-$e$ case and a high-$a$, high-$e$ case. These
	roughly bracket the Planet~9 prior space, though they keep the aphelion
	fixed at RA = $60^\circ$ ($\beta = 62^\circ$) instead of marginalizing over it.
	When a black curve is below the bluest curve, then all pixels at that longitude
	are deep enough to detect it. When the black curve is above the reddest curve,
	then no pixels at that longitude are deep enough to detect it. And when
	the black curve crosses the 70\% quantile (orange), then 70\% of the pixels
	are too shallow to detect it. The top and bottom panels are for a $5 M_\Earth$
	(low: $a = 350 \text{AU}$, $e = 0.1$; high: $a = 550 \text{AU}$, $e = 0.3$)
	and $10 M_\Earth$ (low: $a = 650 \text{AU}$, $e = 0.3$; high: $a = 850 \text{AU}$,
	$e = 0.5$) Planet~9 respectively.
	}
	\label{fig:rlim-ecl}
\end{figure}

\section{Discussion}
As the possible signal curves in Figure~\ref{fig:specs} showed, ACT's
non-detection is not surprising, especially considering that a planet
in an eccentric orbit moves more slowly near aphelion, and is therefore more likely to be located
there. Planet~9's aphelion is predicted to be around $\text{RA} \approx 60^\circ$,
an area where the ACT coverage is quite shallow, corresponding to
a $5M_e$ detection limit of about 350 AU. For comparison, the smallest
expected aphelion distance is a bit less than 400 AU (Table~\ref{tab:prior}).
Hence, at its current depth, ACT can not expect to see Planet~9 if it is
near aphelion.

Because \citetalias{p9-hypothesis} does not provide a well-defined
prior volume, it is hard to quantify what fraction of the Planet~9
parameter space we have probed, but we can make a few simple estimates.
Figure~\ref{fig:rlim-1d} shows the distribution of our distance limits
for the Planet~9-relevant parts of the sky ($|i| < 30^\circ$), and compares
them to the $\sim$ 300--700 AU and $\sim$ 400--1300 AU allowed distance range
for a $5 M_\Earth$ and $10 M_\Earth$ Planet~9 respectively. We probe
about 13\% and 8\% of this distance-position space.
However, that does not take into account the fact that the furthest Planet~9
distances are only expected to occur in some parts of the sky.
The spatial dependence of the predicted Planet~9 distance range
is shown in Figure~\ref{fig:rlim-ecl}, and taking it into account, our
numbers improve to 17\% and 9\% respectively.

The upcoming Simons Observatory (SO) \citep{so_science} will substantially improve on these
bounds. Extrapolating our current results to the expected depth of the combined ACT+SO
data set, we can expect to detect a $5 M_\Earth$ Planet~9 at 500--600 AU near
the expected aphelion location and 500--900 AU over most of the rest of its orbit. This is still not
enough to guarantee a discovery, but it will probe a substantial fraction
of its parameter space. Unlike bounds from optical surveys like Pan-STARRS
and LSST, and even sub-mm ones like WISE, the ACT and SO searches are
only mildly sensitive to Planet~9's physical composition, and are robust to
assumptions about atmospheric emission lines and albedo.

\begin{acknowledgements}
This work was supported by the U.S. National Science Foundation through awards
AST-0408698, AST-0965625, and AST-1440226 for the ACT project, as well as
awards PHY-0355328, PHY-0855887 and PHY-1214379. Funding was also provided by
Princeton University, the University of Pennsylvania, and a Canada Foundation
for Innovation (CFI) award to UBC. ACT operates in the Parque Astron\'omico
Atacama in northern Chile under the auspices of the Comisi\'on Nacional de
Investigaci\'on (CONICYT). Flatiron Institute is supported by the Simons Foundation.

Computations were performed using Princeton Research Computing resources at Princeton University, the Niagara supercomputer at the SciNet HPC Consortium and on the Simons-Popeye cluster of the Flatiron Institute. SciNet is funded by the CFI under the auspices of Compute Canada, the Government of Ontario, the Ontario Research Fund---Research Excellence, and the University of Toronto.

SN thanks Bruce Partridge for extensive comments.
EC acknowledges support from the STFC Ernest Rutherford Fellowship ST/M004856/2 and STFC Consolidated Grant ST/S00033X/1, and from the European Research Council (ERC) under the European Union’s Horizon 2020 research and innovation programme (Grant agreement No.  849169).
Research at Perimeter Institute is supported in part by the Government of Canada through the Department of Innovation, Science and Industry Canada and by the Province of Ontario through the Ministry of Colleges and Universities.
SKC acknowledges support from NSF award AST-2001866.
KMH is supported by NSF through AST 1815887.
NS, DH and AM acknowledge support from NSF grant number AST-1907657.

We gratefully acknowledge the many publicly available software packages that
were essential for parts of this analysis. They include
\texttt{healpy}~\citep{Healpix1}, \texttt{HEALPix}~\citep{Healpix2}, and
\texttt{pixell}\footnote{https://github.com/simonsobs/pixell}.
This research made use of \texttt{Astropy}\footnote{http://www.astropy.org},
a community-developed core Python package for Astronomy \citep{astropy:2013,
astropy:2018}. We also acknowledge use of the
\texttt{matplotlib}~\citep{Hunter:2007} package and the Python Image Library
for producing plots in this paper.

\end{acknowledgements}

\bibliographystyle{act_titles}
\bibliography{refs}

\appendix

\section{Smearing}
\label{sec:smearing}
\subsection{Circular smearing}
Consider a Gaussian beam with standard deviation $\sigma$, such that its profile is
\begin{align}
	b(r) &= e^{-\frac12\frac{r^2}{\sigma^2}} = e^{-\frac12 \frac{x^2+y^2}{\sigma^2}}
\end{align}
where $r = \sqrt{x^2+y^2}$.
Partially uncorrected parallax smears this beam along an ellipse with some semi-major axis $\mu$.
The simplest and worst case of this is smearing along a circle with radius $\mu$, so that's what
we will consider here. This results in the smeared beam
\begin{align}
	b_2(r,\mu) &= \frac{1}{2\pi} \int_0^{2\pi} d\theta e^{-\frac12 \frac{(x-\mu\cos\theta)^2+ (y-\mu\sin\theta)^2}{\sigma^2}}
		\propto b(r) \int_0^{2\pi} d\theta
			e^{\frac{x\mu\cos\theta + y\mu\sin\theta}{\sigma^2}}  \notag \\
		&\approx b(r) \int_0^{2\pi} d\theta
			\left(1 + \frac{x\mu\cos\theta + y\mu\sin\theta}{\sigma^2} + \frac12
			\left(\frac{x\mu\cos\theta + y\mu\sin\theta}{\sigma^2}\right)^2\right) \notag \\
		&\propto b(r)
			\left(1 + \frac14 \frac{r^2\mu^2}{\sigma^4}\right)
\end{align}
where we have assumed $\mu \ll \sigma$ and have ignored any factors that just scale
the overall amplitude of the function. If all values $\mu \in [-\frac\Delta2,\frac\Delta2]$
occur with equal weight, then the average beam across all these will be:
\begin{align}
	b_\text{eff}(r) &= \frac{1}{\Delta} \int_{-\frac{\Delta}{2}}^{\frac\Delta2} d\mu
		b_2(r,\mu) =
		\frac{1}{\Delta} b(r) \int_{-\frac{\Delta}{2}}^{\frac\Delta2} d\mu
		\left(1 + \frac14 \frac{r^2\mu^2}{\sigma^4}\right)
		= b(r)\left(1 + \frac{1}{24}\frac{r^2\Delta^2}{\sigma^4}\right)
	\approx e^{-\frac12\frac{r^2}{\sigma^2}} e^{\frac{1}{24}\frac{r^2\Delta^2}{\sigma^4}}
	= e^{-\frac12 \frac{r^2}{\sigma_\text{eff}^2}}
\end{align}
where
\begin{align}
	\sigma_\text{eff}^2 &= \frac{\sigma^2}{1 - \frac{1}{12}\frac{\Delta^2}{\sigma^2}}
	\approx \sigma^2 + \frac{1}{12}\Delta^2 \quad\Leftrightarrow\quad
	\text{FWHM}_\text{eff}^2 = \text{FWHM}^2 + \frac{2\log(2)}{3} \Delta^2 \label{eq:smear-circ}
\end{align}
So circular smearing adds in quadrature to the beam size.

\subsection{Linear smearing}
We here smear the beam linearly in the x direction with $\mu \in [-\frac\Delta2,\frac\Delta2]$.
\begin{align}
	b_\text{eff} &= \frac{1}{\Delta}\int_{-\frac\Delta2}^\frac{\Delta2} d\mu
		e^{-\frac12 \frac{(x-\mu)^2 + y^2}{\sigma^2}}
	\approx b(r)\int_{-\frac\Delta2}^\frac\Delta2 d\mu
		\left(1 + \frac{x\mu}{\sigma^2} + \frac12 \frac{x^2\mu^2}{\sigma^4}\right)
	\frac{e^{-\frac12\frac{\mu^2}{\sigma^2}}}{\Delta}
	\propto b(r)\left(1 + \frac{1}{12}\frac{x^2\Delta^2}{\sigma^4}\right)
	\approx
	e^{-\frac12\left(\frac{x^2}{\sigma_{x,\text{eff}}} + \frac{y^2}{\sigma^2}\right)}
\end{align}
with $\sigma_{x,\text{eff}}^2 = \sigma^2 + \frac16 \Delta^2$. Since only the x direction was
smeared, the beam is now slightly elliptical. For the purposes of S/N, what matters is
not the shape of the beam, but its area, which has gone from $2\pi\sigma^2$ to
$2\pi\sigma\sigma_{x,\text{eff}}$. We can use this to define an effective overall beam
standard deviation:
\begin{align}
	\sigma_\text{eff}^2 &= \sigma\sigma_{x,\text{eff}} =
	\sigma^2\sqrt{1 + \frac16 \frac{\Delta^2}{\sigma^2}}
	\approx \sigma^2 + \frac{1}{12} \Delta^2
	\quad\Leftrightarrow\quad
	\text{FWHM}_\text{eff}^2 = \text{FWHM}^2 + \frac{2\log(2)}{3} \Delta^2 \label{eq:smear-lin}
\end{align}
Which happens to be the same result as what we got for circular smearing.

\section{Building the detection statistic $z$}
\label{sec:norm-app}
The S/N map produced by the shift-and-stack search is non-Gaussian
with properties that depend on both the distance $r$ and the
position in the map, making it unsuitable as an indication of the
detection strength. However, we found that the following
three-step approach allowed us to transform $S/N$ into a much
more well-behaved detection statistic $z$.

\subsection{Spatial normalization}
\begin{figure}[hpt]
	\centering
	\begin{tabular}{cc}
	Raw & Normalized \\
	\includegraphics[width=0.5\textwidth,trim=0 4mm 7mm 4mm]{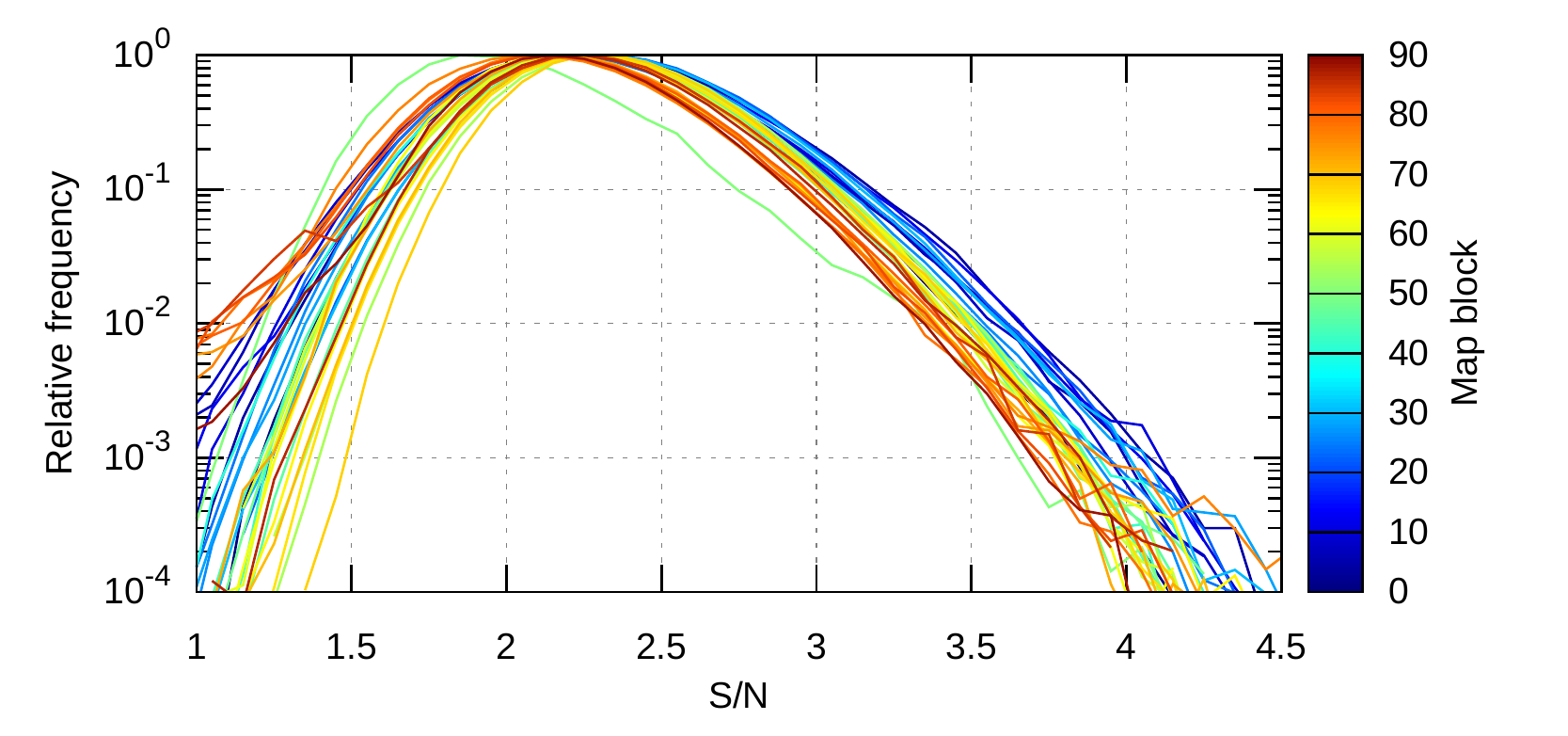} &
	\includegraphics[width=0.5\textwidth,trim=7mm 4mm 0mm 4mm]{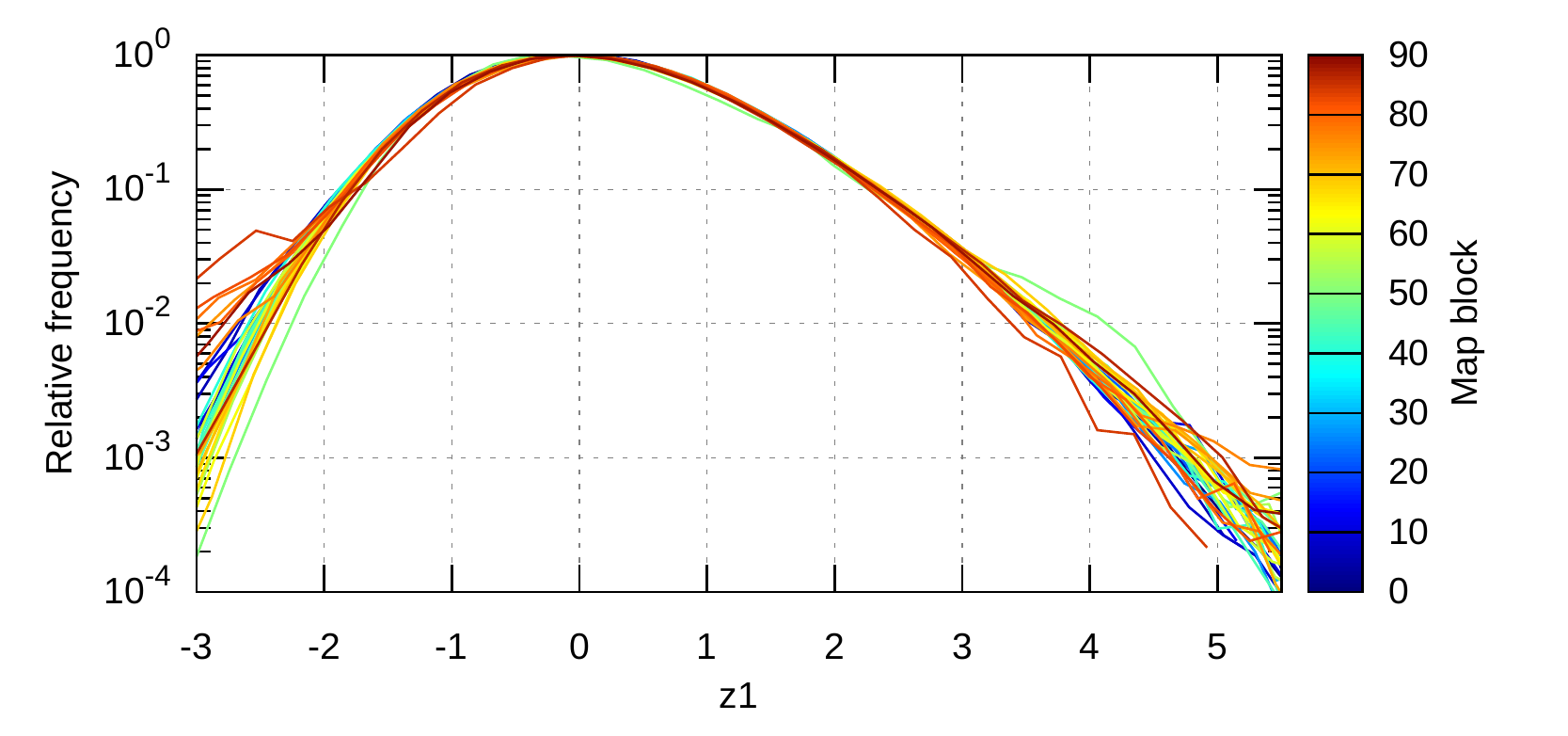}
	\end{tabular}
	\caption{Spatial normalization of the detection statistic.
	\dfn{Left}: Histograms for the shift-and-stack S/N ratio for the 300 AU case.
	Each curve corresponds to a different $5^\circ\times5^\circ$ part of the map. The distribution
	is non-Gaussian and spatially variable. \dfn{Right}: The same histograms after normalizing
	by subtracting the mean and dividing by the standard deviation. After this the distribution
	is no longer spatially variable. The actual spatial normalization used in the search uses
	smaller $0.5^\circ \times 4^\circ$ degree tiles, which are large enough to measure
	the mean and standard deviation reliably, but result in noisier histograms.}
	\label{fig:norm-spat}
\end{figure}

For each $r$ we measure the mean $\mu_\text{spat}(r)$ and
standard deviation $\sigma_\text{spat}(r)$ of the S/N map
as a function of position. \footnote{We do this in $0.5^\circ\times4^\circ$ blocks.
These short-wide blocks were chosen because many features in
the ACT exposure pattern are wider than they are tall. The
block size is a compromise between angular resolution and
sample variance in the estimates.} Using these, we define
the spatially-normalized detection statistic $z_1$ as
\begin{align}
	z_1(r) &= [S/N(r) - \mu_\text{spat}(r)]/\sigma_\text{spat}(r)
\end{align}
 This normalization process is shown in Figure~\ref{fig:norm-spat}.

\subsection{Distance normalization}

\begin{figure*}[htbp]
	\centering
	\begin{tabular}{cc}
		$z_1$ survival function & $z_1$ vs $z^*$ \\
		\includegraphics[height=5cm,trim=8mm 4mm 8mm 6mm]{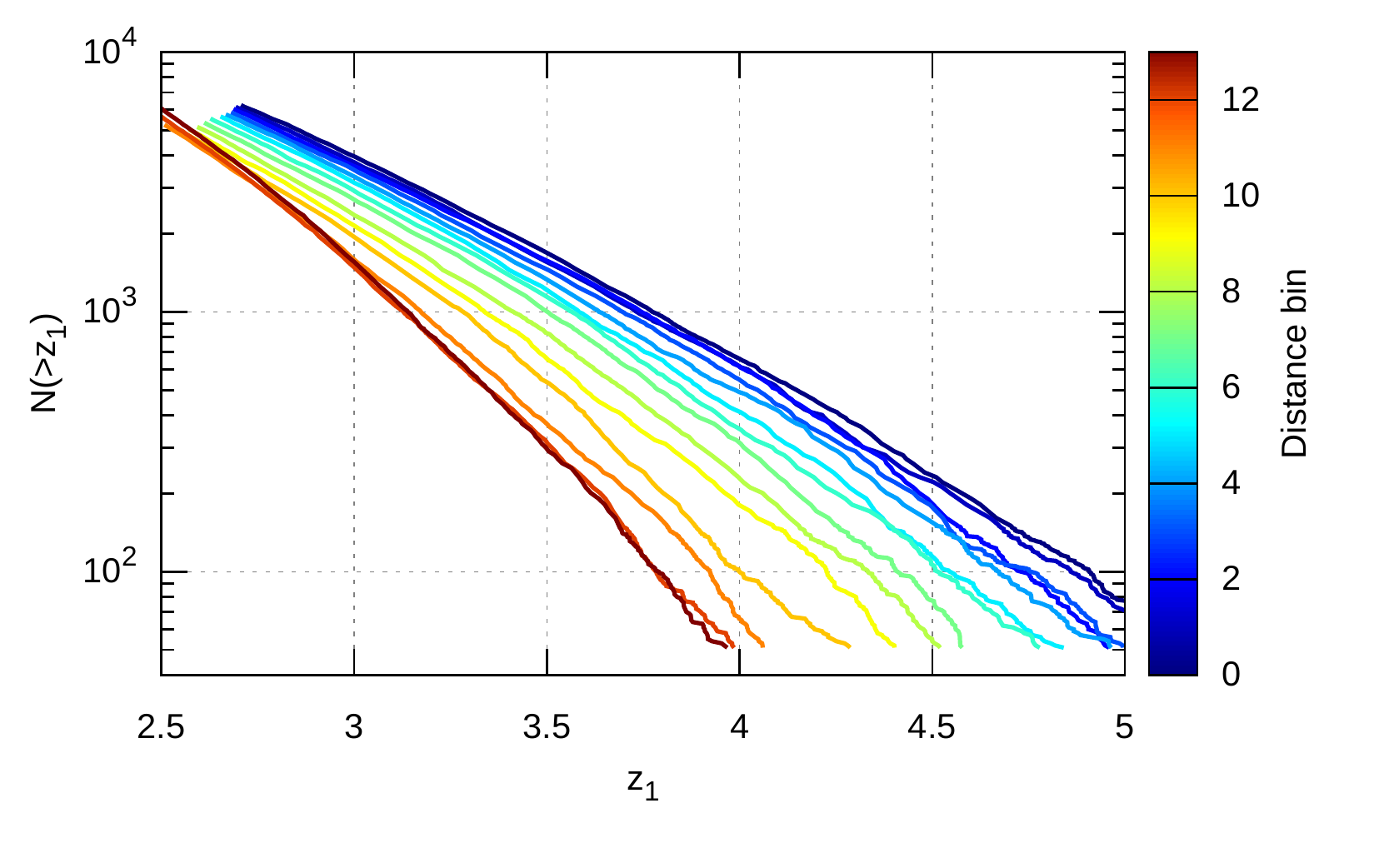} &
		\includegraphics[height=5cm,trim=8mm 4mm 8mm 6mm]{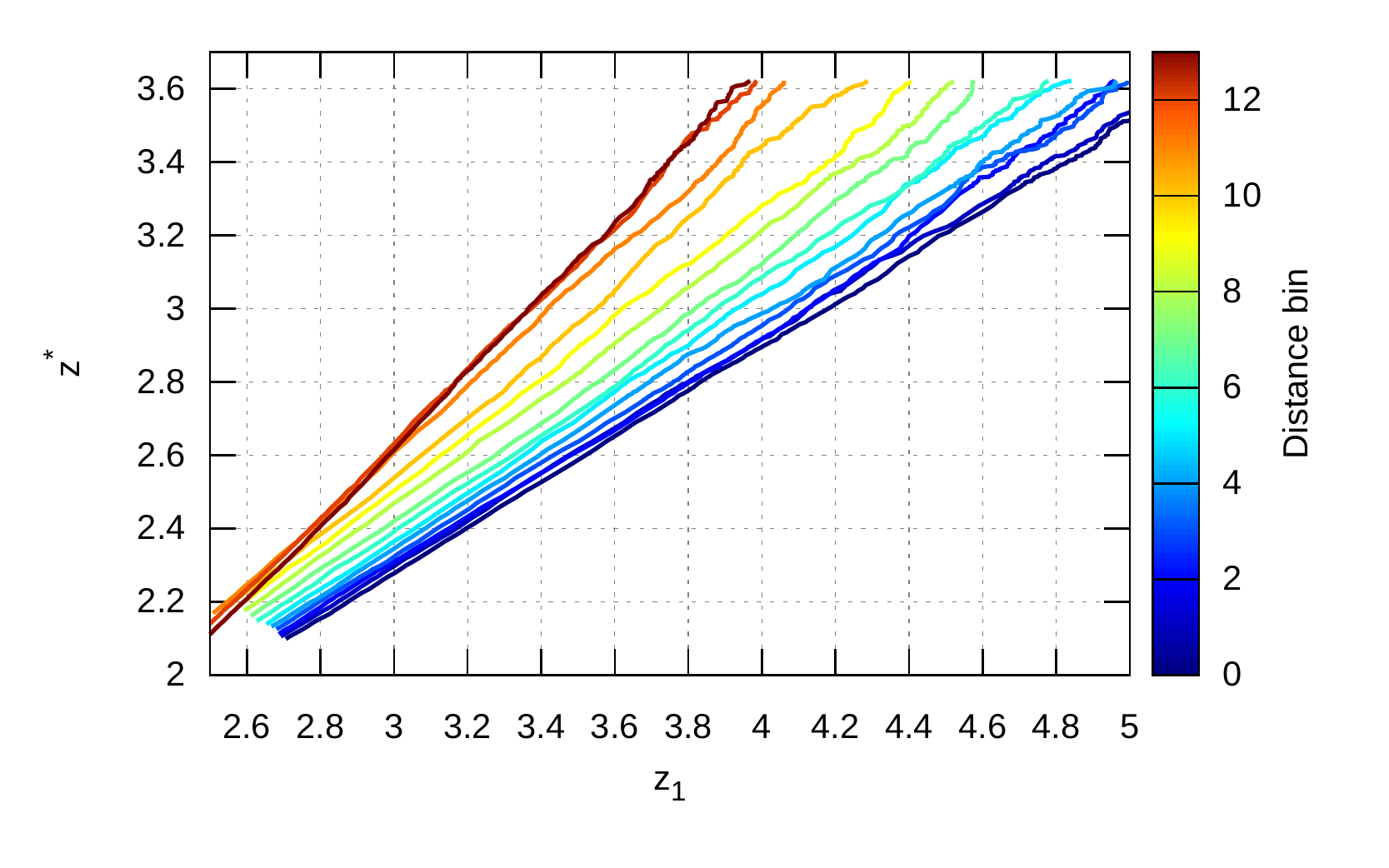} \\
		$z$ survival function & $z$ vs $z^*$ \\
		\includegraphics[height=5cm,trim=8mm 4mm 8mm 6mm]{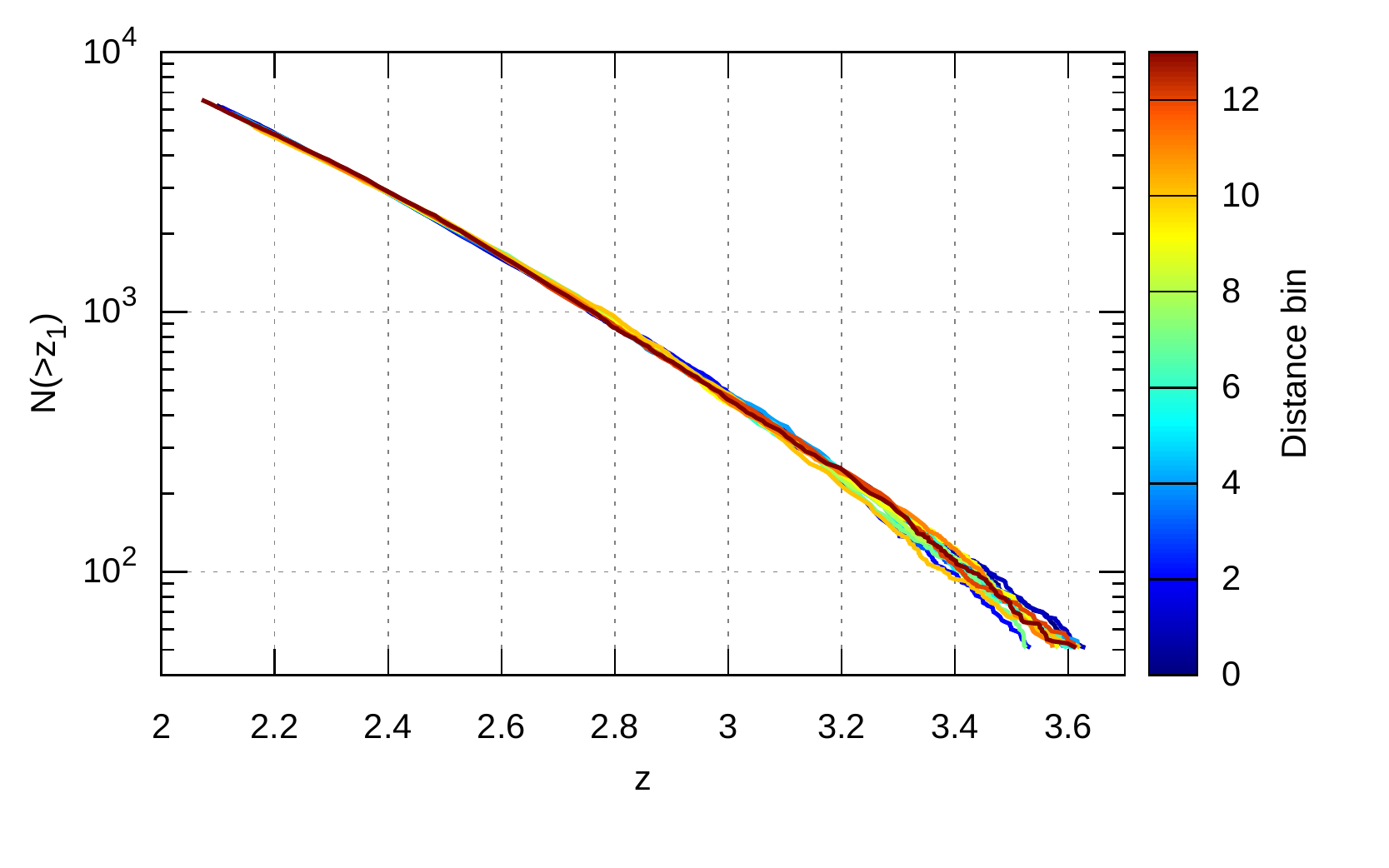} &
		\includegraphics[height=5cm,trim=8mm 4mm 8mm 6mm]{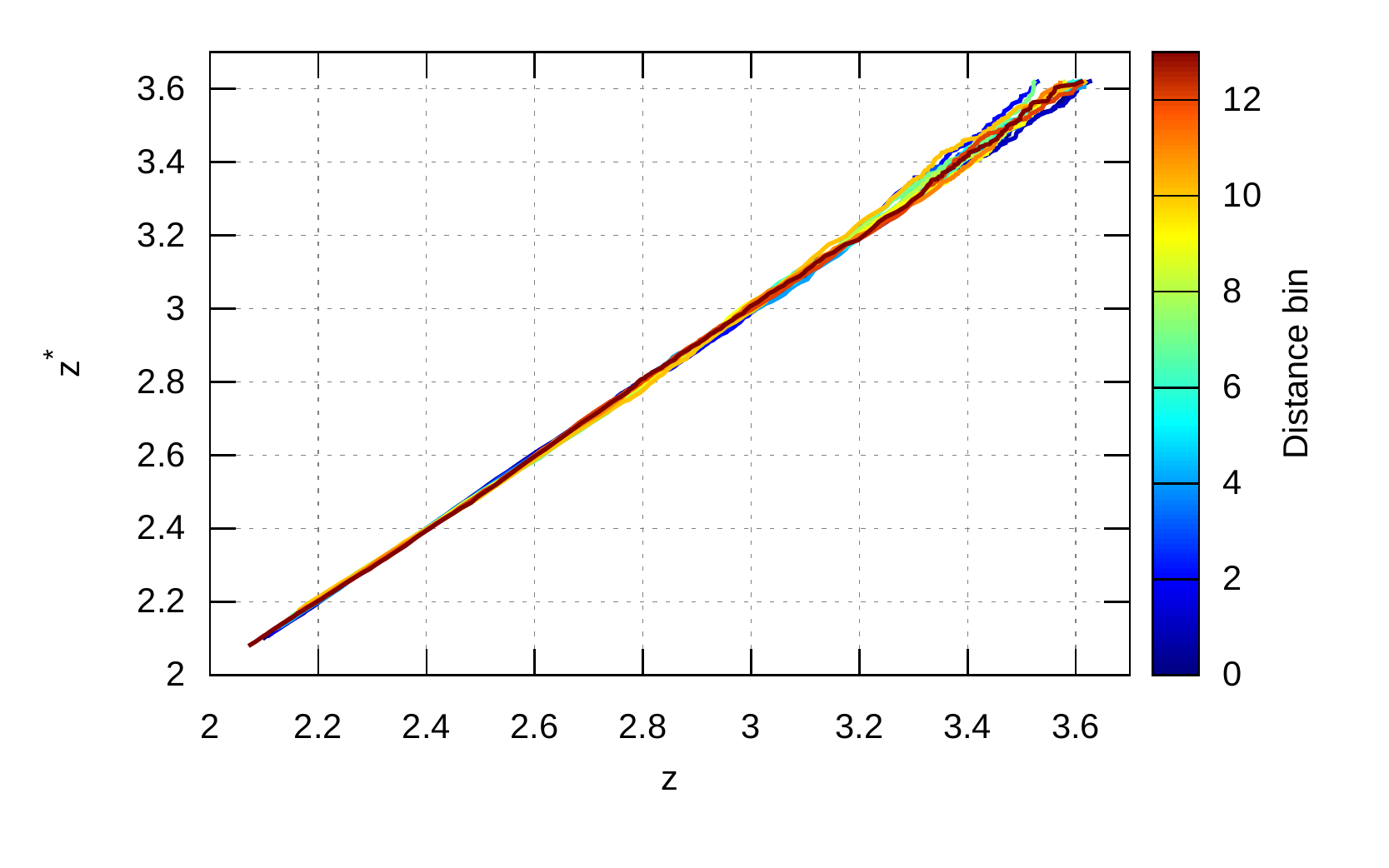}
	\end{tabular}
	\caption{Distance normalization of the detection statistic.
	\dfn{Top left}: The tail of the survival function of $z_1$. Each
	color corresponds to a different distance bin, from 300 AU (blue)
	to 2000 AU (red). The different distances clearly do not follow the
	same distribution. \dfn{Top right}: As top left, but with the survival
	function replaced with the corresponding Gaussian quantile $z^*$.
	For each distance we define $\mu_\text{dist}$ and $\sigma_\text{dist}$
	as the best-fit linear offset and slope of its curve.
	\dfn{Bottom left}: The survival function for the distance-normalized
	detection statistic $z = (z_1 - \mu_\text{dist})/\sigma_\text{dist}$.
	All distances now follow the same distribution.
	\dfn{Bottom right}: $z$ vs. the empirical Gaussian quantiles $z^*$.
	They are practically identical.}
	\label{fig:norm-dist}
\end{figure*}
We then build the empirical survival function
$N(z_1>x)$ for all peaks in the tile with $z_1>1$ and map this to
a corresponding Gaussian quantile
\begin{align}
	z^* = -\sqrt{2}\text{ erf}^{-1}(2N/n-1) \label{eq:zstar}
\end{align}
The value of $n$ controls how far into the tail of
a Gaussian survival function we map our empirical
survival function. Its exact value is not important
as long as $n > \text{max}(N)$. It should be kept
constant for all values of $r$ to ensure that equally
rare values of $z_1$ map to the same $z^*$ for all distances.
We chose $n = A_\text{tile}/A_\text{peak}
\approx (12^\circ/2')^2 = 1.3\cdot10^6$, with $A_\text{tile} = (12^\circ)^2$
being the area of the tile, and $A_\text{peak} = (2')^2$ being the
approximate feature size in the $z_1$ map. In principle eq.~\ref{eq:zstar}
could be used to directly normalize $z_1$, but in practice
there are too few samples in the tail. However, $z^*$ turns
out to be very well approximated as a linear function of $z_1$.\footnote{This means
that the upper tail of the $z_1$ distribution (which is what we care about for
feature detection) is nearly Gaussian, even though the whole distribution isn't.}
We use this to define the fully normalized detection statistic
\begin{align}
	z(r) &= [z_1(r) - \mu_\text{dist}(r)]/\sigma_\text{dist}(r) \label{eq:z}
\end{align}
where $\mu_\text{dist}$ and $\sigma_\text{dist}$ are the
offset and slope of the function $z^*(z_1)$.
This process is illustrated in Figure~\ref{fig:norm-dist}. Inserting the
expression for $z_1$ into equation~\ref{eq:z}, we get the full normalization
\begin{align}
	z(r) &= (S/N(r) - \mu_z(r))/\sigma_z(r) = \xi(S/N)
\end{align}
where we have defined $\mu_z = \mu_\text{spat} + \sigma_\text{spat}\mu_\text{dist}$
and $\sigma_z = \sigma_\text{spat}\sigma_\text{dist}$; and we have implicitly defined
the function $\xi$.

\section{Ad-hoc filter}
\label{sec:ad-hoc-appendix}
The map noise power can be approximately modeled as $1/\beta(\ell,\ell_\text{knee},\alpha)$,
where $\beta$ is the Butterworth filter profile
\begin{align}
	\beta(\ell,\ell_\text{knee},\alpha) = 1/[1 + (\ell/\ell_\text{knee})^\alpha]
\end{align}
This noise power spectrum takes the form of a power law with slope $\alpha \approx -4$
at low $\ell$ (mainly caused by atmospheric emission) which transitions to
a flat ``noise floor'' around the multipole $\ell_\text{knee} \approx 3000$
where photon noise and detector readout noise start to dominate.\footnote{
	$\ell_\text{knee}$ is frequency-dependent, taking values of about 2000/3000/4000
	at f090/f150/f220, but because this issue was discovered after the frequency
	maps had already been combined we will just use a representative 3000 here.}
However, it turned out that $N$, the noise model we use for our time-ordered
data analysis (see eq.~\ref{eq:mapmaking}) does not capture the full correlation
structure of the atmosphere, and ends up underestimating the effective $\ell_\text{knee}$
by a factor of two, i.e. $\ell_\text{knee}' \approx 1500$. This means that
our matched filter did not suppress noise in the $1500\lesssim \ell \lesssim 3000$
range as much as it should be, leading to a loss in S/N.

To correct for this, we replace the map inverse covariance matrix $U^{-1}$
with $\beta(\ell,3000,-4) U^{-1}$. Accordingly $\rho = R^TU^{-1}\hat m$
is remapped at $\rho \rightarrow \beta \rho$. What happens to
$\kappa = \text{diag}(R^TU^{-1}R)$ is harder to estimate. We can approximate
it as $\kappa \rightarrow q\kappa$, where
$q$ a single number representing the weighted average of the extra filter $\beta(\ell,3000,-4)$ over
all multipoles,
\begin{align}
	q = \frac{\sum_\ell (2\ell+1) W(\ell) \beta(\ell,3000,-4)}{\sum_\ell (2\ell+1) W_\ell}
\end{align}
with the weights $W(\ell)$ being a harmonic-space approximation of the original matched filter,
\begin{align}
	W(\ell) &=  \beta(\ell,1500,-4)B(\ell)
\end{align}
where $B(\ell)$ is the beam and $\beta(\ell,1500,-4)$ approximates the original $U^{-1}$.
However, in the end the normalization of $\kappa$ does not matter, since it is absorbed
by the simulation-based debiasing we do to account for the effect of mean sky subtraction
in Section~\ref{sec:bias}.

The ad-hoc filter could have been avoided if we had computed the full $\hat m$ and had
done the full matched filter in pixel-space instead of using the ``rhs'' computational
shortcut described in Section~\ref{sec:skysub}. This is a protential improvement for
future analyses.

\section{ACT data set details}
\label{sec:data-appendix}
Table~\ref{tab:data} summarizes the ACT data sets used in this analysis.

\begin{table*}[htp]
	\centering
	\hspace*{-20mm}\begin{tabular}{llr!{--}rr!{--}rl}
		\toprule
		\bf Survey & \bf Patch & \multicolumn{2}{c}{\bf RA ($^\circ$)} & \multicolumn{2}{c}{\bf dec ($^\circ$)} & {\bf Data sets} \\
		\midrule
		ACT DR4 & D1  & 140 & 161 &  -5 &   6 & PA1 2013 \\
		ACT DR4 & D5  & -19 &  13 &  -7 &   6 & PA1 2013 \\
		ACT DR4 & D6  &  19 &  48 & -11 &   1 & PA1 2013 \\
		ACT DR4 & D56 & -23 &  54 & -10 &   7 & PA1+PA2 2014--2015, PA3 2015 \\
		ACT DR4 & D8  & -12 &  18 & -52 & -32 & PA1+PA2+PA3 2015 \\
		ACT DR4 & BN  & 102 & 257 &  -7 &  22 & PA1+PA2+PA3 2015 \\
		ACT DR4 & AA  &   0 & 360 & -62 &  22 & PA2+PA3 2016 \\
		AdvACT  & AA  &   0 & 360 & -62 &  22 & PA4+PA5+PA6 2017--2019 \\
		ACT day & BN     & 102 & 257 &  -7 &  22 & PA1+PA2 2014--2015, PA3 2015 \\
		ACT day & Day-N   & 162 & 258 &   3 &  20 & PA2+PA3 2016, PA4+PA5+PA6 2017--2019 \\
		ACT day & Day-S   & -25 &  60 & -52 & -29 & PA4+PA5+PA6 2017--2019\\
	\bottomrule
	\end{tabular}
	\caption{ACT data sets used in the analysis. They are identical to those used
	in ACT DR5 \citep{act-coadd-2020}, except for the inclusion of data from the 2019
	observing season, and the exclusion of Planck (too low resolution) and
	ACT MBAC (too low sky coverage). PA\{1-6\} refers to the individual detector arrays
	in the instrument, with PA\{1,2\} covering the f150 band, PA\{3,5,6\} covering
	f090 and f150, and PA4 covering f150 and f220.}
	\label{tab:data}
\end{table*}

\end{document}